\documentclass[journal, twoside]{IEEEtran}
\usepackage{graphicx}
 \usepackage{cuted}
\usepackage{nicefrac}
\usepackage{cite}
\usepackage{amsmath}
\usepackage{subscript}
\usepackage{trfsigns}
\usepackage{float}
\usepackage{soul}
\usepackage{xcolor}
    \if@twocolumn
        \newenvironment{widetext}
            {%
                \begin{strip}
                \rule{\dimexpr(0.5\textwidth-0.5\columnsep-0.4pt)}{0.4pt}%
                \rule{0.4pt}{6pt}
                \par 
                \parindent \@parindent
            }%
            {%
                \par
                \hfill\rule[-6pt]{0.4pt}{6.4pt}%
                \rule{\dimexpr(0.5\textwidth-0.5\columnsep-1pt)}{0.4pt}
                \end{strip}
            }
    \else
        \newenvironment{widetext}{}{}
    \fi

\ifCLASSINFOpdf
\else
\fi
\begin{document}
%
\title{Electrodynamics of the Superconducting State \\in Ultra-Thin Films at THz Frequencies}
%
%
%
\author{Uwe~S.~Pracht, Eric~Heintze, Conrad Clauss, Daniel~Hafner, Roman~Bek, David~Werner, Sergey~Gelhorn, Marc~Scheffler, Martin~Dressel, Daniel Sherman, Boris Gorshunov, Konstantin S.~Il\'{}in, Dagmar~Henrich, and Michael~Siegel
\thanks{U. S. Pracht, E. Heintze, C. Clauss, D. Hafner, R. Bek, D. Werner, S. Gelhorn, M. Scheffler, and M. Dressel are with 1. Physikalisches Institut Universit\"at Stuttgart, Pfaffenwaldring 57, D-70550 Stuttgart. Germany} 
\thanks{Daniel Sherman is with Faculty of Exact Sciences, Bar-Ilan University Ramat-Gan, 52900 Israel}
\thanks{{Boris Gorshunov is with A.M. Prokhorov Institute of General Physics, Russian Academy of Sciences, Moscow 119991, Russia, and
Moscow Institute of Physics and Technology (State University), Dolgoprudny, Moscow Region 141700, Russia}
}
\thanks{K. Il\'{}in. D. Henrich, and M. Siegel are with Institut f\"ur Mikro- und Nanoelektronische Systeme (IMS), Karlsruher Institut f\"ur Technologie (KIT), Hertzstra\ss{}e 16, D - 76187 Karlsruhe, Germany}
\thanks{Manuscript submitted January 29, 2013, revised manuscript submitted March 20, 2013}}
%
%
\markboth{IEEE Transactions on THz Science and Technology, 2013}%
{Pracht \MakeLowercase{\textit{et al.}}: Electrodynamics of the Superconducting State in Ultra-Thin Films at THz Frequencies}
%
\maketitle
\begin{abstract}
We report on {terahertz} frequency-domain spectroscopy (THz-FDS) experiments in which we measure charge carrier dynamics and excitations of thin-film superconducting systems at low temperatures in the THz spectral range. The characteristics of the set-up and the experimental procedures are described comprehensively. We discuss the single-particle density of states and a theory of electrodynamic absorption and optical conductivity of conventional superconductors. We present the experimental performance of the setup at low temperatures for a broad spectral range from 3 to 38\,cm\textsuperscript{-1} (0.1 - 1.1\,THz) by the example of ultra-thin films of weakly disordered superconductors niobium nitride (NbN) and tantalum nitride (TaN) with different values of critical temperatures $\textit{T\textsubscript{c}}$. Furthermore, we analyze and interpret our experimental data within the framework of conventional Bardeen-Cooper-Schrieffer (BCS) theory of superconductivity. By and large, we find the properties of our NbN and TaN thin films to be well described by the theory. Our results on NbN resemble tendencies towards anomalous behavior of the ratio $\mathbf{2\Delta(0)/\textit{k}_\textit{B}\textit{T}_\textit{c}}$ as a function of $\textit{T\textsubscript{c}}$. 
\end{abstract}
\begin{IEEEkeywords}
Frequency-domain THz spectroscopy, superconducting thin films, BCS theory, density of states of a superconductor, optical conductivity of superconductors, superconductor-insulator transition, TaN, NbN.
\end{IEEEkeywords}
%
\IEEEpeerreviewmaketitle
\section{Introduction}
\IEEEPARstart{T}{hin} films of superconductors are of both technological and academic interest. Properties of interacting electron systems are governed by the electronic density of states (DOS) which changes upon reducing spatial dimensions. Thus, dimensional reduction 3D$\rightarrow$2D may affect optical, electronic, and thermodynamic properties of the system. In superconductivity, studying quasi-2D systems turned out to be a particularly fruitful field and led to remarkable findings such as insulators featuring a superconducting gap \cite{She12}, pseudogap in conventional s-wave superconductors \cite{bat10}, or extremely strong-coupling superconductivity in Kondo lattices \cite{Miz11}. Besides pure academic interest, thin-film superconductors play a key role in many state-of-the-art applications such as SQUIDs, thermal and electrical switches using Josephson junctions, or microwave resonators \cite{vis12}. Ultra-thin films of niobium nitride (NbN) and tantalum nitride (TaN) studied in this work are suited for a novel approach to build single-photon detectors \cite{Eng12a,Hen12,Ker06,Gol01,Zha03,Nat12}. Apart from applications, NbN has recently gained attention as a model system for the superconductor-insulator transition (SIT): Tunneling studies \cite{Cho09} reveal a gapped DOS slightly above $T_c$ in highly disordered films as it has been found for titanium nitride (TiN) films \cite{bat10}. This peculiar state resembles the well-established pseudogap in high-$T_c$ cuprate superconductors. In the case of NbN, however, $T_c$ is about one order of magnitude higher compared to TiN. Thus, measurements on NbN in the vicinity of the SIT are much less demanding than in the case of TiN \cite{Pra12,Dri12}. Consequently, results obtained for NbN might serve as a key to understand the still puzzling pseudogap state in high-$T_c$ cuprates. 

THz time-resolved spectroscopy \cite{Bec11}, THz frequency-domain spectroscopy (THz-FDS) \cite{Gor93}, THz pump-THz probe spectroscopy \cite{Mat12}, far-IR laser spectroscopy \cite{Tes10, Tes11, Sin10}, and far-IR Fourier-spectroscopy \cite{Dre91} measurements on comparably thick films of NbN have been performed previously in a comparable spectral range. With our approach, however, we push experimental limits towards even lower frequencies and temperatures which allow us to study films with a higher degree of disorder and thus a smaller energy gap \cite{Cho09}.
THz spectroscopy in the range from 0.03 to 1.5\,THz (1 to 50\,cm$^{-1}$ in the units common for spectroscopists) has proved to be a powerful tool to study many questions among thin film superconductors \cite{dre08}, since the superconducting energy gap $2\Delta$ of many compounds happens to fall in the corresponding energy range of the order of a millielectronvolt ($\sim 0.25$\,THz). Importantly, no contacts or surface structuring are required which might affect the sample\'{}s properties, or complicate the analysis because of contact effects.
The setup described below is a quasi-optical Mach-Zehnder interferometer that enables us to measure amplitude and phaseshift of THz radiation passing through a thin film on a plane-parallel substrate. From these experimental data, any complex optical function such as complex conductivity $\hat{\sigma}=\sigma_1+\mathrm{i}\sigma_2$ or the complex permittivity $\hat{\epsilon}=\epsilon_1+\mathrm{i}\epsilon_2$ as well as free-electron parameters e.g., scattering rate $\nicefrac{1}{\tau}$ or plasma frequency $\omega_p$, can be calculated. 

This review paper is structured as follows. In section \ref{sec:theo}, we give a brief introduction in the BCS theory of superconductivity focusing on the single-particle density of states and the electromagnetic absorption and the optical conductivity. Furthermore, in section \ref{sec:experimental} we describe our experimental techniques and the analysis approaches employed. Section \ref{sec:exp} features experimental data on thin film superconductors NbN and TaN, which not only play an interesting role in current research but also nicely demonstrate the agreement between theory and experiment.
\section{\label{sec:theo}Electromagnetic Absorption and Optical Conductivity of a Superconductor}
In 1956, Cooper proposed an instability of a metal\'{}s electronic structure towards the condensation of free electrons into electron-electron pairs in the presence of a (arbitrary small) positive electron-electron net interaction \cite{Coo56} and, by this, laid the foundation for BCS theory of superconductivity \cite{Bar57}. By the exchange of a virtual phonon at sufficiently low temperatures, the repulsive Coulomb force between two (itinerant) electrons may be overcompensated lowering the energy of each electron by $\Delta(T)$, and lead to a bound electron-electron state, a Cooper pair, with the total binding energy $2\Delta(T)$. The phonon-mediated interaction is most effective in case of electrons with opposite momenta $\mathbf{k}_1=\mathbf{k}$ and $\mathbf{k}_2=-\mathbf{k}$ and antiparallel spins $s_1=\nicefrac{1}{2}$ and $s_2=-\nicefrac{1}{2}$. Consequently, a Cooper pair has no net momentum and zero spin, thus obeying Bose-Einstein statistics. All Cooper pairs are allowed to occupy the same state which leads to the formation of a macroscopic phase-coherent condensate. Since the Cooper-pair binding energy typically is of the order of a few millielectronvolts, the bound state is easily destroyed by thermal energy, and thus superconductivity commonly is limited to very low temperatures below a critical temperature $T_c$ (Note: In what follows we will focus on conventional superconductors originating from metallic normal states. For reviews on unconventional superconductivity see e.g., \cite{Basov2005, Ste11, Nor11}).
\subsection{\label{sec:DOS}Single-particle density of states}
\begin{figure}
\noindent \begin{centering}
\includegraphics[scale=0.45]{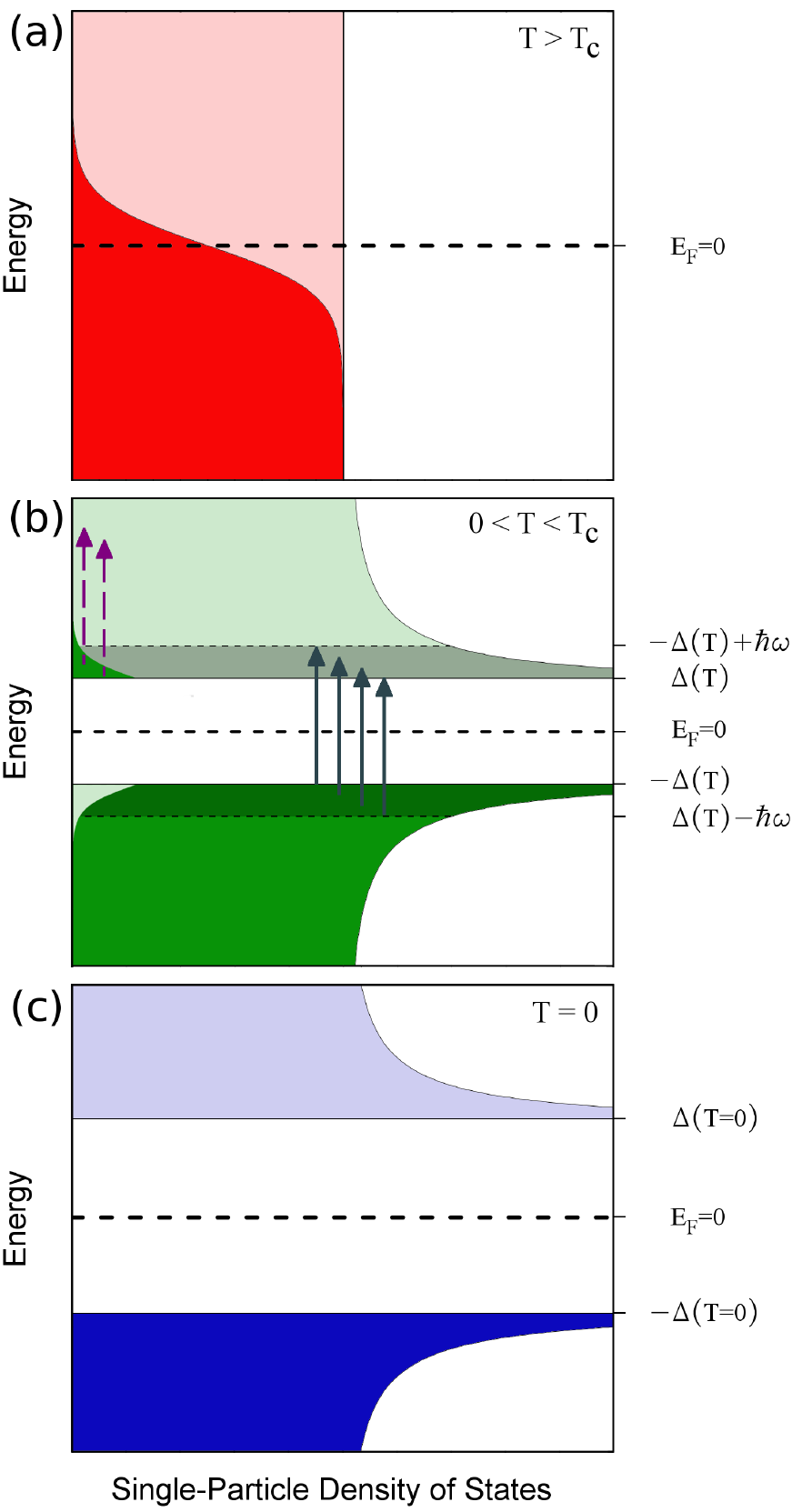}
\par\end{centering}
\caption{\label{fig:DOS}(Color online) Single-particle density states around the Fermi energy $E_F=0$ of a superconductor. (a) At $T>T_c$, the energy states are occupied according to Fermi-Dirac distribution leading to thermal population of some states above and some empty states below $E_F$. (b) At $0<T<T_c$, a temperature-dependent energy gap $2\Delta(T)$ opens around $E_F$ and the DOS diverges at the edges of the gap. Due to finite temperatures, some states above the gap are occupied by thermally activated unpaired electrons. By absorption of photons of arbitrary energy, these thermal electrons contribute to $\sigma_1$. These transitions are marked by dashed arrows and correspond to integral \textit{A} in (\ref{eq:sigma}). Electrons below the gap exclusively contribute for photon energies greater than $2\Delta(T)$. These transitions are marked by solid arrows and correspond to integral \textit{B} in (\ref{eq:sigma}). (c) At $T=0$, no thermal electrons are apparent and $2\Delta(T)$ reaches it\'{}s maximum. }
\end{figure}
The superconducting state features a number of intriguing properties like non-dissipative transport current, perfect diamagnetism, and a peculiar single-particle density of states depending on temperature. At low temperatures in the superconducting state, the DOS of a metal features an energy gap of the size $2\Delta$ around the Fermi energy $E_F$ and diverges at $E_F\pm \Delta$, as shown in Fig. \ref{fig:DOS}. The particular shape of the DOS sensitively depends on temperature. In the normal state, $T>T_c$, the DOS around $E_F$ can be considered constant within the energy range of interest. The energy states are occupied according to Fermi-Dirac distribution around the $E_F$, as shown in Fig. \ref{fig:DOS}(a). Because of the finite temperature, some states above $E_F$ are occupied with thermal electrons while some states below $E_F$ are empty. Upon cooling below $T_c$, the superconducting state evolves and an energy gap $2\Delta(T)$ starts to open symmetrically around $E_F$ relocating the states in the vicinity of $E_F$,as shown in Fig. \ref{fig:DOS}(b). The states which happen to fall in the energy gap are shifted above $E_F+\Delta(T)$ and below $E_F-\Delta(T)$ increasing the DOS at the edges of the energy gap. At $E_F\pm\Delta(T)$, the DOS diverges and becomes infinite. The electrons which occupied the vanished states are bound into Cooper pairs and disappear from the single particle DOS when condensing to a mutual ground state. (In some models suited for e.g., tunneling phenomena, Cooper pairs are considered to be at $E_F$. But we employ a semiconductor-like picture, where Cooper pairs are not apparent). The majority of unpaired electrons remains below the gap, not contributing to the electrodynamic properties. A small number, however, is thermally activated to states above $E_F+\Delta(T)$. These thermally activated unpaired electrons (in the following just thermal electrons) play a key role for electrodynamics at finite temperatures. As the temperature approaches zero, $2\Delta(T)$ rises to it\'{}s maximum and the number of thermal electrons decreases to zero, as shown in Fig. \ref{fig:DOS}(c). (In real systems, the DOS does not actually diverge but is smeared out due to the finite lifetime of the thermal electrons \cite{Sol72}.)
\subsection{\label{sec:Sigma}Electromagnetic properties and optical conductivity}
The gapped and diverging DOS gives rise to significant changes in the electromagnetic absorption and the optical conductivity for frequencies of the same order as the gap frequency $\omega_g=2\Delta/\hbar$, with $\hbar=h/2\pi$ the reduced Planck constant. Considering typical values of the energy gap, these changes are likely to be found in the far-IR, THz and microwave spectral ranges. The optical conductivity of a superconductor was first derived by Mattis and Bardeen \cite{Mat58} and holds for superconductors without significant electron scattering, i.e. where the mean free path $\ell\propto \tau$, with $\tau$ the relaxation time, by far exceeds the superconducting coherence length $\xi$ and penetration depth $\lambda$. The more general formalism for arbitrary values of $\tau$ is first presented in \cite{Zim91}, and casts into the equation (with $E_F\equiv0$)
\begin{equation}
\hat{\sigma}\left(\omega\right)=\mathrm{i}\frac{\sigma_{dc}}{2\omega\tau}\times\left(\underbrace{\int_{\Delta(T)}^{\hbar\omega+\Delta(T)}I_{1}\mathrm{d}E}_{\mathrm{Integral\,A}}+\underbrace{ \int_{\Delta(T)}^{\infty}I_{2}\mathrm{d}E}_{\mathrm{Integral\, B}}\right)\label{eq:sigma}
\end{equation}
with
\\
\begin{widetext}
\[
I_{1}=\tanh\frac{E}{2k_{B}T}\left\{ \left[1-\frac{\Delta(T)^{2}+E\left(E-\hbar\omega\right)}{P_{3}P_{2}}\right]\frac{1}{P_{3}+P_{2}+\mathrm{i}\hbar\tau^{-1}}-\left[1+\frac{\Delta(T)^{2}+E\left(E-\hbar\omega\right)}{P_{3}P_{2}}\right]\frac{1}{P_{3}-P_{2}+\mathrm{i}\hbar\tau^{-1}}\right\}
\]
\begin{eqnarray}
I_{2}=\tanh\frac{E+\hbar\omega}{2k_{B}T}\left\{ \left[1+\frac{\Delta(T)^{2}+E\left(E+\hbar\omega\right)}{P_{1}P_{2}}\right]\frac{1}{P_{1}-P_{2}+\mathrm{i}\hbar\tau^{-1}}-\left[1-\frac{\Delta(T)^{2}+E\left(E+\hbar\omega\right)}{P_{1}P_{2}}\right]\frac{1}{-P_{1}-P_{2}+\mathrm{i}\hbar\tau^{-1}}\right\} \nonumber\\ +\tanh\frac{E}{2k_{B}T}\left\{ \left[1-\frac{\Delta(T)^{2}+E\left(E+\hbar\omega\right)}{P_{1}P_{2}}\right]\frac{1}{P_{1}+P_{2}+\mathrm{i}\hbar\tau^{-1}}-\left[1+\frac{\Delta(T)^{2}+E\left(E+\hbar\omega\right)}{P_{1}P_{2}}\right]\frac{1}{P_{1}-P_{2}+\mathrm{i}\hbar\tau^{-1}}\right\} \nonumber
\end{eqnarray}
\end{widetext}
\[
P_{1}=\sqrt{\left(E+\hbar\omega\right)^{2}-\Delta(T)^{2}}\hspace{2cm}P_{2}=\sqrt{E^{2}-\Delta(T)^{2}}\hspace{2cm}P_{3}=\sqrt{\left(E-\hbar\omega\right)^{2}-\Delta(T)^{2}}
\]
and $k_B$ the Boltzmann constant. (For numeric calculations, a more convenient version of (\ref{eq:sigma}) is given in \cite{Zim91}.) Real and imaginary parts of the complex conductivity follow from (\ref{eq:sigma}) with $\sigma_1=\text{Re}(\hat{\sigma})$ and $\sigma_2=\text{Im}(\hat{\sigma})$. 

\begin{figure}[h]
\noindent \begin{centering}
\includegraphics[scale=0.40]{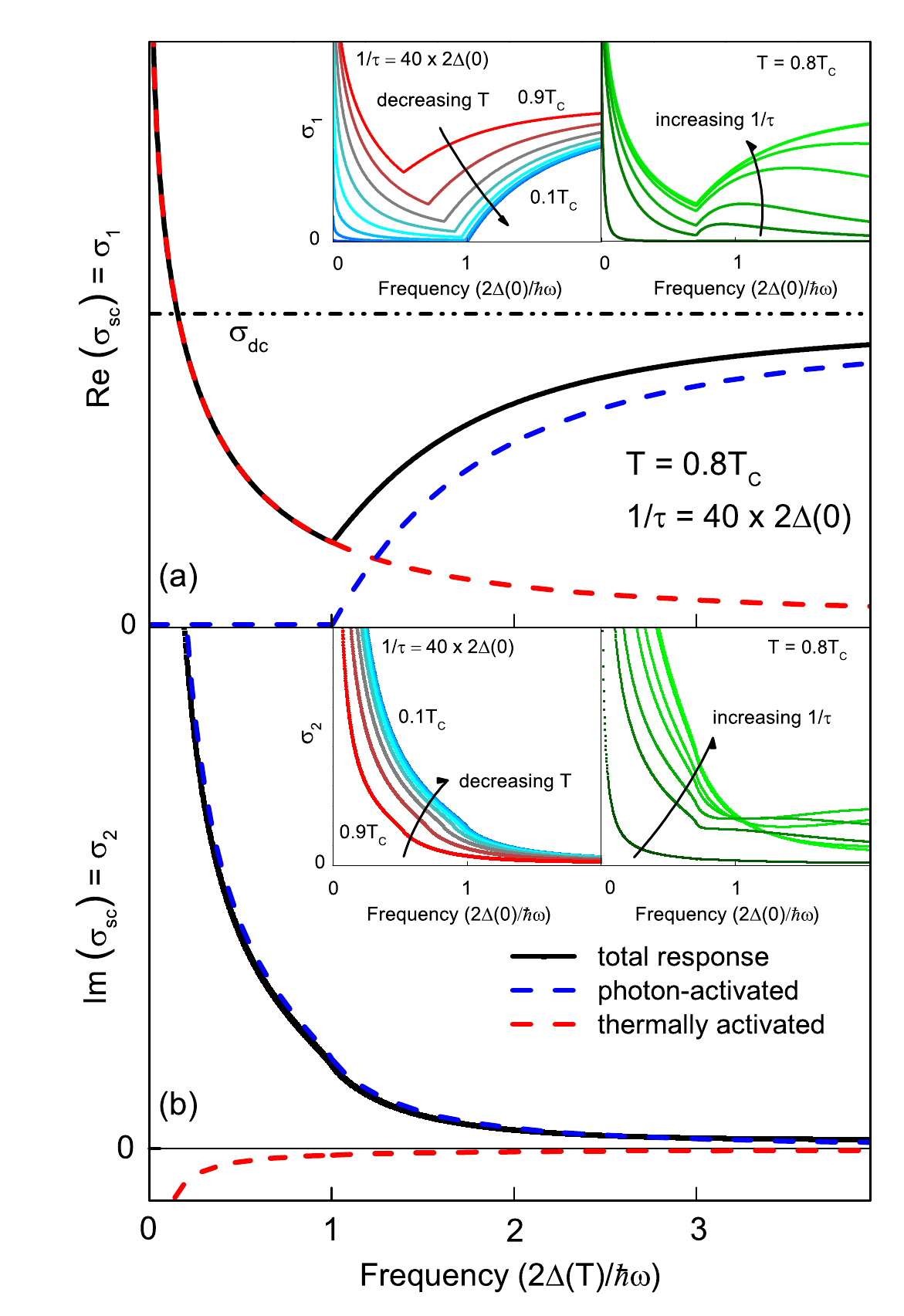}
\par\end{centering}
\caption{\label{fig:sigma_theo}(Color online)(a) Real and (b) imaginary parts of the complex optical conductivity (solid line) of a superconductor as a function of frequency based on (\ref{eq:sigma}) for $T=0.8T_c$. Dashed lines disentangle the contributions of both the thermal electrons (Integral \textit{B} in (\ref{eq:sigma}) and photon-activated electrons (Integral \textit{A} in (\ref{eq:sigma}) to the conductivity. Insets of (a) and (b) show $\sigma_1$ and $\sigma_2$ for different temperatures ($T/T_c$ = 0.9, 0.8, 0.7, 0.6, 0.5, 0.4, 0.3, 0.2, and 0.1) and scattering rates ($\nicefrac{1}{\tau}$ = 40, 20, 4, 2, 0.8, 0.4, and 0.04 $\times 2\Delta(0))$, respectively. (Used parameters $2\Delta(0)=25$\,cm$^{-1}$, $T_c=10$\,K, $\sigma_{dc}=1066$\,$\Omega^{-1}$cm$^{-1}$)}
\end{figure}

The first integral $\int_{\Delta(T)}^{\hbar\omega+\Delta(T)}I_1\mathrm{d}E$ in (\ref{eq:sigma}) (in the following labeled integral $A$) accounts for the electromagnetic response of the Cooper pairs, while the second one $\int_{\Delta(T)}^{\infty}I_2\mathrm{d}E$ (integral $B$) accounts for the thermal electrons. Typical graphs for $\sigma_1(\omega)$ and $\sigma_2(\omega)$ calculated using (\ref{eq:sigma}) are shown in Fig. \ref{fig:sigma_theo}. Besides the total response taking into account both integrals $A$ and $B$, Fig. \ref{fig:sigma_theo} disentangles the contributions resulting from Cooper pairs and thermal electrons. For finite temperatures below $T_c$ and frequencies higher than $\omega_g$, electrons are activated across the energy gap by absorbing photons, as shown in Fig. \ref{fig:DOS}b), thus giving rise to a finite $\sigma_1$ (This process corresponds to the breakup of a Cooper pair, adding two \textit{quasiparticles} - a quantum superposition of electron and hole wavefunctions e.g., \cite{Ann11} - to the single-particle DOS). With increasing photon energy, the range from where electrons may be excited across the gap becomes greater enabling enhanced absorption and consequently rising $\sigma_1$. The contribution of thermal electrons is finite, however, inferior to the previously discussed contribution. This is mainly due to the comparably small amount of thermal electrons and the small DOS at energies higher than $\Delta(T)+\hbar\omega_g$. (Note: integral $B$ covers the energy range $[\Delta(T),\infty)$ since thermal electrons occupy states up to arbitrary high energies. Integral $A$ covers the energy range $[\Delta(T),\hbar\omega+\Delta(T)]$, however, concerning $\sigma_1$ it yields non-zero values only up to $\hbar\omega-\Delta(T)$.) For frequencies smaller than $\omega_g$, however, the photon energy becomes insufficient to excite electrons across the gap, and consequently the electromagnetic response is completely governed by intraband transitions of thermal electrons. The number of possible transitions by photon absorption increases with decreasing frequency due to the diverging DOS. Consequently, $\sigma_1$ rises strongly as the frequency approaches zero. At a certain frequency, $\sigma_1$ exceeds the normal state conductivity $\sigma_{dc}$. This becomes obvious from spectral weight arguments, since the area beneath $\sigma_1(\omega)$ is constant for all temperatures according to Ferrell-Tinkham-Glover sum rule \cite{Fer58, Tin59}. The missing area beneath $\sigma_1(\omega)$ in the superconducting state is relocated to both the $\delta$-function at zero frequency and to the low-frequency range, where $\sigma_1$ rises above the normal state conductivity. For $\omega= 0$, the contribution of thermal electrons merges with the infinitely high $\delta$-like conductivity
\rule{\dimexpr(0.5\textwidth-0.5\columnsep-1pt)}{0.4pt}%
\rule[-10pt]{0.4pt}{10.4pt}
\\
\\
\\
\\
\\
\\
\\
\\
\\
\\
\\
\\
\rule{\dimexpr(0.5\textwidth-0.5\columnsep-0.4pt)}{0.4pt}%
\rule{0.4pt}{10pt}
of the superconducting condensate at zero frequency. (Note: equation (\ref{eq:sigma}) does not include the $\delta$-function contribution to $\sigma_1$.)\\ 
While the real part $\sigma_1$ of the optical conductivity features both considerable contributions from integrals $A$ and $B$, the finite imaginary part $\sigma_2$ is mainly due to the superconducting condensate (at sufficiently low temperatures and frequencies), whose conductivity is derived via the 1. London equation \cite{dre02}
\begin{eqnarray}
\frac{\partial j_s(\omega, t)}{\partial t}=\frac{n_s e^2}{m}E(\omega,t)\nonumber\\
\nonumber\\
\longrightarrow j_s(\omega, t)=\frac{n_se^2}{m}\int_{-\infty}^{t} E(\omega, t^{\prime})\mathrm{d}t^{\prime} 
\label{eq:london}
\end{eqnarray}
(with $n_s$ the superfluid density, $e$ the elementary charge, and $m$ the electron mass) relating the time derivative of the (local) supercurrent density $j_s(\omega, t)$ to an external perturbation $E(\omega,t)$. In case of an electromagnetic wave, $E(\omega, t)=E_0 \exp(-\mathrm{i}\omega t)$, (\ref{eq:london}) then becomes
\begin{equation} 
j_s(\omega, t)=\frac{\mathrm{i}}{\omega}\frac{n_s e^2}{m}E(\omega, t) \label{eq:london2}
\end{equation}
Equation (\ref{eq:london2}) has the same structure as Ohm\'{}s law, and thus we interpret the proportionality factor $\frac{\mathrm{i}}{\omega}\frac{n_s e^2}{m}$ between $j_s(\omega, t)$ and $E(\omega,t)$ as the conductivity of the superconducting condensate, which is purely imaginary and thus only contributes to $\sigma_2$. Upon increasing frequency, $\sigma_2$ decays as $\nicefrac{1}{\omega}$. Consequently, the strongest contribution is at energies well below $2\Delta(T)$, much lower than addressed in our study, i.e. at microwave frequencies \cite{Tur03,Ste08}. Kramers-Kronig analysis underlines the above qualitative picture yielding the same result \cite{dre02}.
Integral $A$ yields finite values up to $\hbar\omega+\Delta(T)$ which calls for the integration range given in (\ref{eq:sigma}). For frequencies larger than $\omega_g$, Cooper pairs are destroyed, and thus the overall out-of-phase response changes. This is signaled by a weak kink of $\sigma_2$ at the energy gap, as shown in Fig. \ref{fig:sigma_theo}. In analogy to an electric LC circuit, the frequency dependence of $\sigma_2$ is often interpreted in terms of frequency dependent kinetic inductance \cite{Maz02,Mes69}. The impedance of the circuit rises towards higher frequencies leading to a reduced conductivity. A finite scattering rate $\nicefrac{1}{\tau}$ accounted for in (\ref{eq:sigma}) further influences the behavior of the complex conductivity. Upon increasing $1/\tau$ (going from clean-limit to dirty-limit superconductors), the rise in $\sigma_1$ for $\omega > \omega_g$ is gradually enhanced, and $\sigma_2$ features a strongly pronounced kink at $\omega_g$ for scattering rates comparable to the energy gap, see insets of Fig. \ref{fig:sigma_theo}(a) and (b). 
\section{\label{sec:experimental}Experiment and data processing}
\subsection{\label{sec:Thz}THz frequency-domain spectroscopy}
\begin{figure}[h]
\noindent \begin{centering}
\includegraphics[scale=0.22]{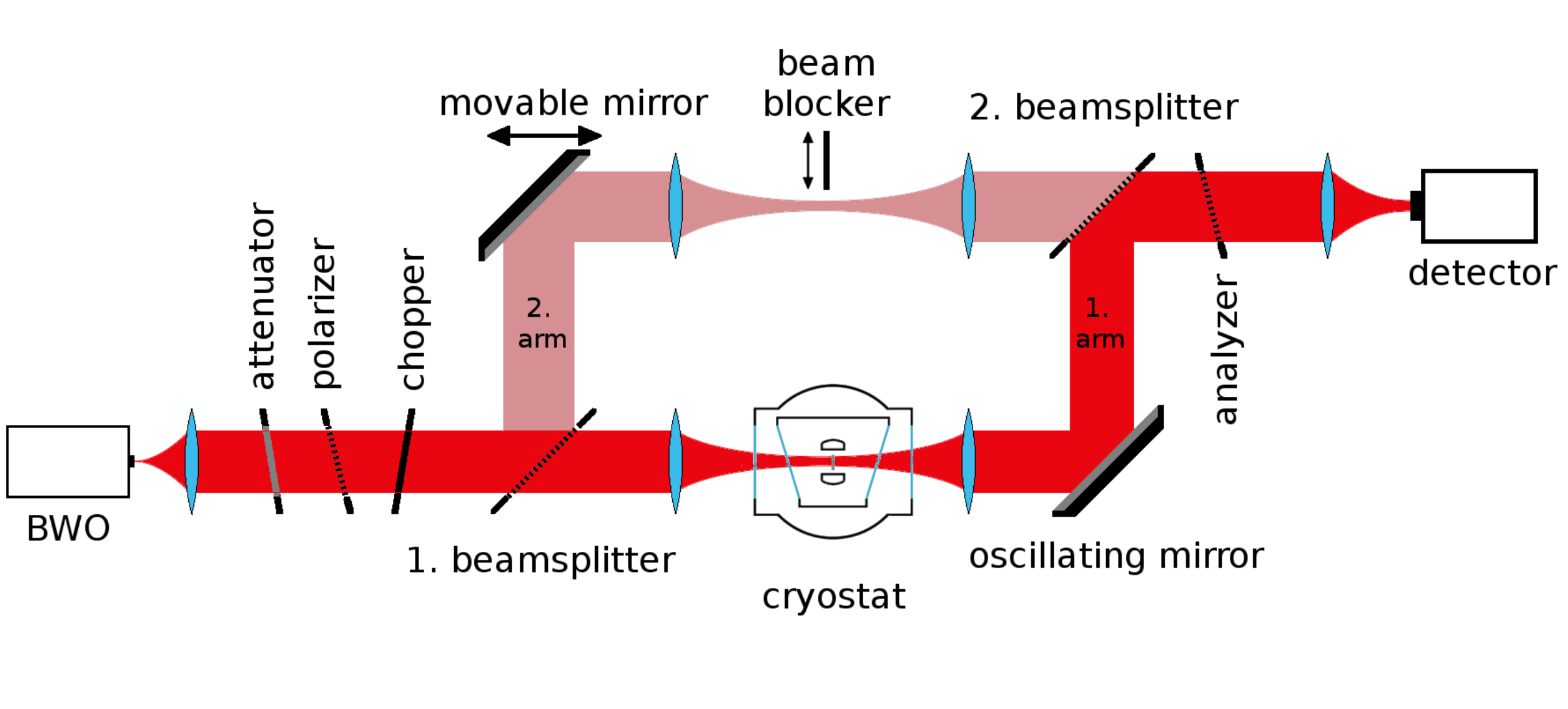} \\
\includegraphics[scale=0.29]{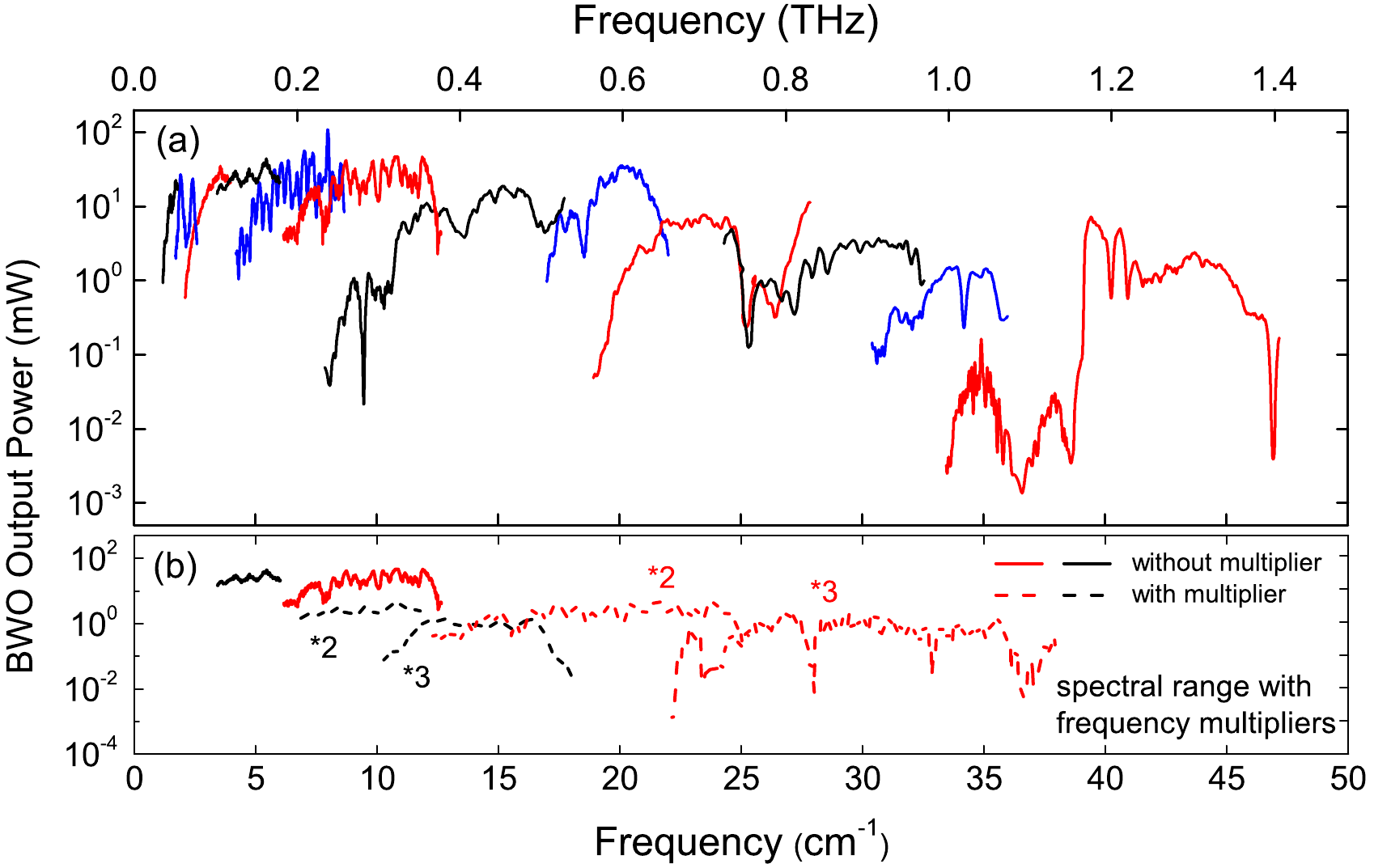}
\par\end{centering}
\caption{\label{fig:machzehnder}(Color online) Top: Simplified sketch of the utilized
Mach-Zehnder interferometer. Attenuators, polarizer, chopper, and analyzer are tilted with respect to the beam propagation axis to suppress standing waves between these parts. Measuring the phaseshift requires both arms of the interferometer, whereas the second arm is blocked for transmission measurements. Bottom: (a) BWO output power of several sources versus frequency. The range from 1 to 47\,cm$^{-1}$ is covered entirely. (b) The available spectral range of a source can be increased using passive frequency doubler and tripler (dashed lines). The signal intensity in the doubled/tripled spectral range, however, is 3 to 4 orders of magnitude smaller and may require use of a $^4$He-cooled bolometer for detection.}
\end{figure}

Our THz studies are based on the frequency-domain spectroscopy technique. Coherent, monochromatic and continuous THz radiation is generated by frequency-tunable backward-wave-oscillators (BWO) that provide frequencies from 1 to 47\,cm$^{-1}$ \cite{Koz98}. This spectral range is covered by a number of different BWO radiation sources (in the following just \textit{sources}), see bottom of Fig. \ref{fig:machzehnder}. The power of the radiation strongly depends on the
frequency and may be as large as 100\,mW for low-frequency and significantly smaller (0.1\,mW)
for high-frequency sources. As a consequence, the low-frequency radiation power might be too \textit{strong} and requires use of attenuators to avoid sample heating or detector overload. The great benefit of FDS compared to time-domain spectroscopy (TDS) is to process frequencies step by step with a stability and resolution exceeding $\Delta\omega/\omega=10^{-6}$ depending on the BWO power supply. 

While TDS excites all available energies simultaneously and has to struggle with minor signal strength, FDS can directly probe narrow-band excitations and resolve detailed line shapes. The high output power leads to high signal-to-noise ratio that can reach values up to $10^6$. Frequency doublers and triplers (Virginia Diodes) that allow extending the frequency range of a separate source can effectively be used. Loss of the radiation intensity is easily 
compensated by use of a $^4$He bolometer (see below). After the radiation is generated by the BWO, it is collimated by a Teflon, quartz, or polyethylene lens and guided by aluminum mirrors and wire grids, see top of Fig. \ref{fig:machzehnder}, which act as beam splitters depending on polarization. Other lenses are used to focus the radiation on the sample under study and the detector. Attenuators, polarizer, chopper, and analyzer are tilted with respect to the beam propagation axis to suppress standing waves between these parts. The quasi-optical part of the experiment is a Mach-Zehnder interferometer which allows us to measure both amplitude and phaseshift of THz radiation passing through a sample.
\begin{figure}
\noindent \begin{centering}
\includegraphics[scale=0.66]{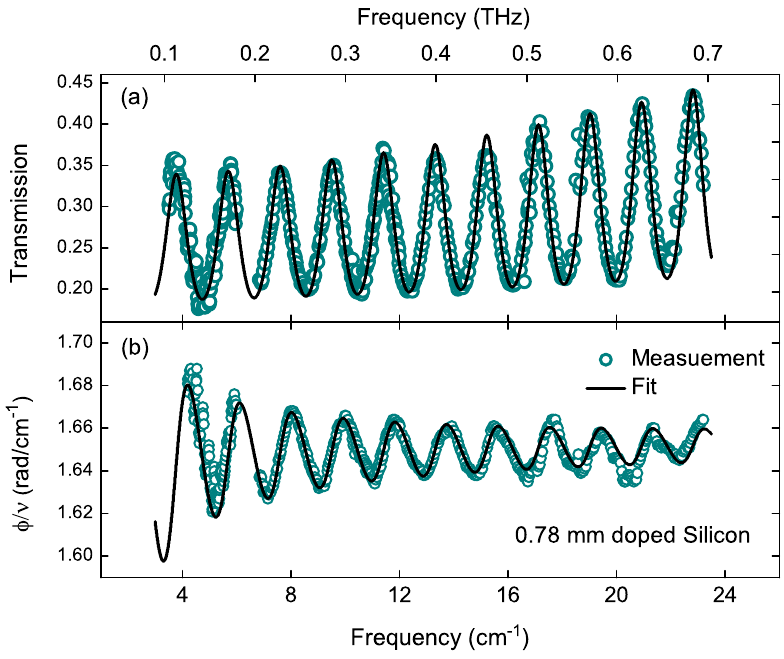}
\par\end{centering}
\caption{\label{fig:fit-vs-data}(Color online) Typical raw data of (a) transmission and (b) relative phaseshift versus frequency together with simultaneously performed fits according to (\ref{eq:Tr}) and (\ref{eq:drude4}). The sample is a $\sim0.78$\,mm thick substrate of weakly doped silicon at room temperature. Note: the phaseshift is the measured quantity, however, it is common to work with phaseshift divided by frequency. This will be done throughout the entire work.}
\end{figure}
The radiation is detected by either a Golay cell or a $^{4}$He-cooled bolometer (Infrared Laboratories), depending on signal strength. For transmission measurements, the radiation is mechanically chopped to make it detectable using a lock-in technique.

Determining the spectra of optical parameters of the plane-parallel sample involves measurements of the transmission and phaseshift spectra from which the spectra of real and imaginary parts of complex permittivity, conductivity, etc., are calculated directly and without use of any additional analysis (such as Kramers-Kronig analysis).

Recording a transmission spectrum is performed with the second arm of the interferometer blocked. It consists of two separate steps. First, the signal intensity versus frequency is recorded with no sample in the beam path. This transmission spectrum is regarded 100\% transmission and used as measurement calibration. Second, the transmission spectrum is recorded with the sample in the beam path and the absolute transmission spectrum is obtained as a result of division of the two corresponding data arrays. In this way we can account for the frequency-dependent output power of the radiation passing through the measurement channel. Typical data (300 K) of a 0.78 mm thick substrate of weakly doped silicon ($n=3.5\times10^{14}$\,cm$^{-3}$) are shown in Fig. \ref{fig:fit-vs-data}(a) \cite{Pra12}. The oscillation pattern is due to multiple reflections inside the sample which acts as a Fabry-P\'erot resonator.

To measure phaseshift spectra, the second arm has to be unblocked. Again, the measurement consists of two separate steps: calibration without and measurement with the sample in the beam path. In order to obtain reliable data, one has to guarantee optimal interference of the two beams at the analyzer. This is provided by aligning the spectrometer elements and achieving maximal ratio between minimal (destructive interference) and maximal (constructive interference) signal strength. The interference signal is detected using a lock-in technique. However, the signal is modulated not using the chopper (amplitude modulation) but the oscillating mirror (phase modulation), as shown in Fig. \ref{fig:machzehnder}. During a measurement process, the movable mirror is automatically kept in a position corresponding to zero-order destructive interference (minimal signal) while the frequency range is swept. The mirror position versus frequency is recorded when the sample is out of (calibration scan) and in (measurement) the optical path, and the phaseshift spectrum is calculated from the difference between the two spectra. Typical phaseshift data taken on the same 0.78 mm thick substrate of doped silicon is provided in Fig. \ref{fig:fit-vs-data}(b). (Note: the measured quantity is \textit{phaseshift} $\phi$ (rad), however, in this paper we will always refer to \textit{relative phaseshift} $\phi/\nu$ (rad/cm$^{-1}$) with $\nu=\omega/(2\pi c $) and $c$ the speed of light in vacuum). 

Optical anisotropy is an issue that has to be taken into account carefully. E.g., the Fabry-P\'erot resonances can become more complex \cite{Sch09}. Many common substrates for thin-film deposition are birefringent such as Al$_2$O$_3$ or NdGaO$_3$. To align the polarization direction \textbf{E} of the radiation to sample\'{}s axes of interest, one can adjust either the sample or the polarization angle of the 1. beam splitter (followed by corresponding turning of the 2. beam splitter, setting it to 90$^{\circ}$ relative to the 1. beam splitter). In case of optical anisotropy of both substrate and film and misaligned optical axes, the optical set-up and analysis has to be customized \cite{Ost11}. 
\subsection{\label{analysis single}Data analysis of single-layer systems}
Transmission and phaseshift data are fitted by well-known Fresnel equations for multiple reflections at media boundaries. For a single layer system, the transmission $\mathcal{T}$
and phaseshift $\phi$ read \cite{dre02}
\begin{equation}
\mathcal{T} = 
\frac{\left(\left(1-R\right)^{2}+4R\sin^{2}\varphi\right)\exp\{-\alpha
d\}}{\left(1-R\exp\{-\alpha d\}\right)^{2}+4R\exp\{-\alpha
d\}\sin^{2}\left(\beta+\varphi\right)} \label{eq:Tr} \\
\end{equation}
\\
\begin{eqnarray}
\phi = \frac{2\pi
nd}{\lambda_{0}}-\arctan\left(\frac{k\left(n^{2}+k^{2}-1\right)}{\left(k^{2}+n^{
2}\right)\left(2+n\right)n}\right)+\nonumber \\
\arctan\left(\frac{R\exp\{-\alpha
d\}\sin^{2}\left(\beta+\varphi\right)}{1-R\exp\{-\alpha
d\}\cos^{2}\left(\beta+\varphi\right)}\right)\label{eq:Ph}
\end{eqnarray}
\\
where $R=(1-2n+n^2+k^2)(1+2n+n^2+k^2)^{-1}$ is the reflectivity, 
$\varphi=\arctan\{-2k(1-n^2-k^2)^{-1}\}$ is the phase change upon reflection at an interface between the two media,
$\beta=2\pi n d /\lambda_0$ is the angle by which the phase of the radiation is changed upon
travelling through a medium of thickness $d$ for a given wavelength $\lambda_0$ and $\alpha=4\pi k /\lambda_0$ is
the power absorption coefficient. The quantities $n,k$ are real and imaginary parts of
the complex refractive index $\sqrt{\hat{\epsilon}}=\hat{n}=n+\mathrm{i}k$ in the given case of a non-magnetic material.
Equations (\ref{eq:Tr}) and (\ref{eq:Ph}) do not account for any frequency dependence for the optical constants $n$ and $k$, which typically depend on frequency and reveal the properties of interest in this study. To include resonant absorption (e.g. Lorentz oscillators) or conductivity (Drude behavior), corresponding terms are used. For example, to describe the free carrier response within the Drude conductivity model we can introduce complex permittivity $\hat{\epsilon}=\epsilon_1+\mathrm{i}\epsilon_2$,
and e.g., in case of a simple metal
the Drude formula \cite{dre02} 
\begin{eqnarray}
\epsilon_{1}(\omega)=\epsilon_{\infty}-\frac{\omega_{p}^{2}}{\omega^{2}+\tau^{-2}} \label{eq:drude1}\\\nonumber\\
\epsilon_{2}(\omega)=\frac{1}{\omega\tau}\frac{\omega_{p}^{2}}{\omega^{2}+\tau^{-2}}\label{eq:drude2}
\end{eqnarray}
or rewritten in terms of the complex optical conductivity $\hat{\sigma}(\omega)=\mathrm{i}\frac{\omega}{4\pi}(\epsilon_{\infty}-\hat{\epsilon}(\omega))$
\begin{eqnarray}
\sigma_1(\omega) =\sigma_{dc}\frac{1}{1+\omega^2\tau^2}\label{eq:drude3}\\\nonumber\\
\sigma_2(\omega) =\sigma_{dc}\frac{\omega \tau}{1+\omega^2\tau^2}\label{eq:drude4}
\end{eqnarray}
where $\omega_p$ is the plasma frequency, $\sigma_{dc}=\nicefrac{\omega_P^2 \tau}{4\pi}$ is the dc conductivity, $\epsilon_{\infty}$ is the value of the dielectric constant for frequencies much higher than addressed in this study, and $\nicefrac{1}{\tau}$ is the relaxation rate. The solid lines in Fig. \ref{fig:fit-vs-data} are fits based upon (\ref{eq:Tr}) and (\ref{eq:Ph}). The fitting parameters $n$ and $k$ are expressed in terms of $\epsilon_1,\epsilon_2,\sigma_1$ and $\sigma_2$ using well-known expressions, \cite{dre02}, and, by this, the frequency dependence according to Drude theory enters the transmission and phaseshift fits via (\ref{eq:drude1})-(\ref{eq:drude4}). Clearly, this yields an excellent description of the experimental data, i.e. this silicon substrate is a text book representative of optical Drude behavior \cite{Gri90, Dre06, nas01}. 

Another analysis approach is to fit the Fabry-P\'erot transmission peaks separately, with two advantages: first, the noise is least pronounced in the vicinity of the Fabry-P\'erot maxima, and the largest sensitivity is reached here. Second, this routine provides more freedom in the case of materials with unknown properties where no explicit function for frequency dependence is known. To do so, each Fabry-P\'erot maximum at some given frequency $\omega_0$ is fitted separately and yields a set of material parameters at $\omega_0$. Processing each Fabry-P\'erot transmission peak in this way eventually yields the desired frequency-dependent optical functions. 
\subsection{\label{analysis multi}Data analysis of multi-layer systems}
Equations (\ref{eq:Tr}) and (\ref{eq:Ph}) hold for a single layer. For a two (or more)- layer
system, however, another boundary has to be accounted for. To disentangle properties of the two media (e.g. film on a substrate) the two experimental quantities (transmission and phaseshift) have to be measured. The response functions of the film are calculated from the two corresponding spectra. This requires detailed knowledge of the substrate parameters, which have to be determined beforehand in a separate experiment. The Fresnel equations are extended by terms describing another boundary reflection and are found elsewhere e.g., \cite{dre02} or \cite{Bor05}. To improve the interaction of probing radiation with the material under study, a resonant Fabry-P\'erot technique has been developed where two specimens of the thin film under study are mounted face-to-face realizing a resonant cavity \cite{Zap13}. This technique is useful in case of superconducting films, where it can be difficult to obtain the dynamical conductivity $\sigma_1$ with sufficient accuracy due to huge positive values of $\sigma_2$. Fabrication of such a resonant cavity, however, remains a challenging task \cite{Zap13}.
\subsection{\label{sec:cryo}Optical cryostats for THz spectroscopy and standing-wave problem}
\begin{figure}
\noindent \begin{centering}
\includegraphics[scale=0.40]{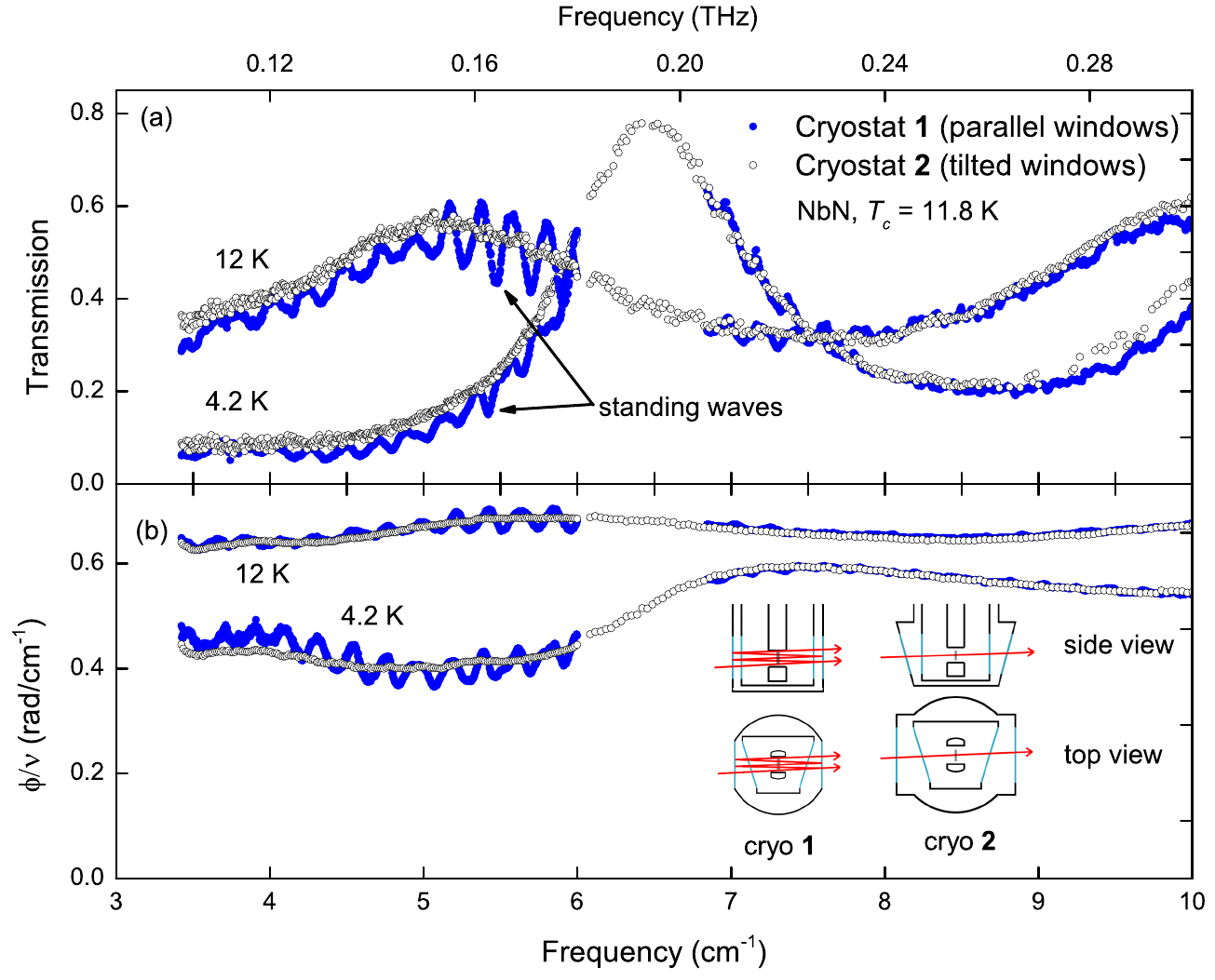}
\par\end{centering}
\caption{\label{fig:standingwaves}(Color online) Raw transmission (a) and relative phaseshift (b) versus frequency measured with parallel-outer-window cryostat "1" (filled circles) and tilted-outer-window cryostat "2" (open circles) at two temperatures 4.2 and 12\,K (sample: NbN, $T_c=11.8$\,K). Towards low frequencies, standing waves are strongly pronounced for cryostat "1". With increasing frequency the standing waves become less pronounced. Simplified sketches of the window arrangement in top- and side-view positions of both cryostats are shown in (b).}
\end{figure}
To access low temperatures we employ two different home-built $^4$He-bath cryostats labeled "1" and "2" here. The most notable difference among these cryostats is the arrangement of the optical inner and outer windows. Diffraction and standing waves distorting the measured spectra become a severe problem especially at low frequencies below $10\,$cm$^{-1}$, since the THz radiation wavelength becomes of the same order as typical setup dimensions, e.g. the distances between inner windows of the cryostat, polarizers, lenses, etc.. To reduce this problem, the windows are made of thin Mylar ($\sim20$\,$\mu m$) foils which are slightly curved due to inner vacuum. The outer windows of cryostat "1" are oriented perpendicularly to the beam still leading to interference which results in multiple reflections and noticeable standing waves. Cryostat "2" was designed with both tilted inner and outer windows (tilting angle $\sim$10$^{\circ}$), see inset of Fig. \ref{fig:standingwaves}(b). Transmission and relative phaseshift spectra taken with both cryostat "1" and "2" are displayed in Fig. \ref{fig:standingwaves} for comparison. Clearly, standing waves are much less pronounced on cryostat "2". Obviously, the tilted windows have to be larger than the non-tilted ones to not reduce the beam cross section. Consequently, the amount of thermal radiation entering the sample chamber is bigger which introduces additional heating, complicating thermal control at temperatures below the boiling point of $^4$He at ambient pressure. 

Another common approach to reduce the standing wave pattern in the measured spectra is modulating the electron-acceleration voltage (0.5 to 6\,kV) applied to the BWO, i.e. modulating the frequency of the THz radiation making it less monochromatic. Narrow-band features of the spectra might be washed out by this frequency averaging, and therefore the high-voltage-modulation amplitude $V_{p-p}$ (0 to 50\,V) has to be adjusted to find an optimum between reduced standing waves and frequency fidelity.
\section{\label{sec:exp}Experiment}
\begin{figure}
\noindent \begin{centering}
\includegraphics[scale=0.35]{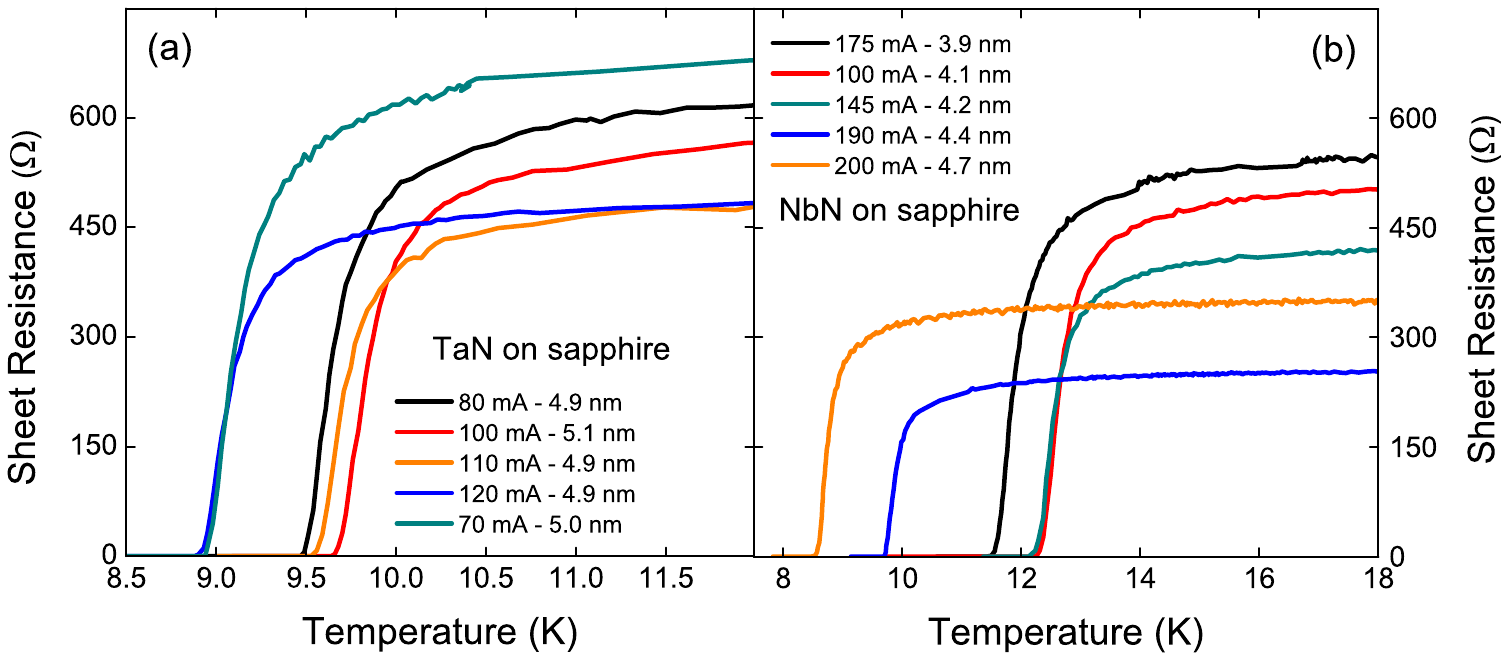}
\par\end{centering}
\caption{\label{fig:rsq}(Color online) Sheet dc resistance versus temperature for different films of (a) TaN and (b) NbN. Different values of $T_c$ are obtained for different deposition currents while sputtering. We estimate the Ioffe-Regel parameter $k_F\ell$ to vary between 4.5 and 5.5 for the NbN films under study far off the superconductor-insulator transition.}
\end{figure}
The films under study were deposited on 10\,mm $\times$ 10\,mm R-plane-cut birefringent sapphire substrates using reactive dc-magnetron sputtering. The NbN and TaN films have a thickness of $d$\,=\,3.9 to 4.7\,nm and 4.9 to 5.1\,nm, respectively. Since the thickness $d$ is similar to the superconducting coherence length $\xi$ \cite{Eng12a}, \cite{Sem09} and much smaller than the magnetic penetration depth, the films are considered ultra-thin, i.e. quasi-2D, from the superconducting point of view. However, the electron mean free path $\ell$ is about one order of magnitude smaller than $d$ \cite{Eng12a}, \cite{Sem09}, which justifies the 3D expressions used in our  analysis. Depending on the deposition current $I_d$ while sputtering, the critical temperature is lowered in the case of NbN from 12.2\,K ($I_d=100$\,mA) to 8.5\,K (200\,mA) and in case of TaN from 9.7\,K (70\,mA) to 8.6\,K (120\,mA), as shown in Fig. \ref{fig:rsq}. Comparison with NbN film deposited in the identical way \cite{Sem09} and \cite{Cho09}, allows us to estimate the amount of disorder via the Ioffe-Regel parameter $k_F\ell$ (with $k_F$ the Fermi wave vector) that varies between 4.5 and 5.5 for our NbN films. Thus, we consider our samples to be only weakly disordered and far off the SIT. For more details on the samples and deposition process see \cite{Ill12} and \cite{Hen12}. We investigate several NbN and TaN samples at temperatures above and below $T_c$ in the frequency range 3.4 to 38\,cm$^{-1}$. To cover this frequency range, 5 different sources were used. Starting at low frequencies, the output power of the first 4 sources was high enough to be detected by a Golay cell. In contrast, to achieve a reasonable signal-to-noise ratio with the high-frequency source, a $^4$He-cooled bolometer was necessary. Transmission and phaseshift measurements were carried out on both main axes of the birefringent sapphire substrate. The samples were mounted into the optical cryostats in such a way that radiation hits the substrate first and subsequently the film. 
\section{Results}
\begin{figure}
\noindent \begin{centering}
\includegraphics[scale=0.81]{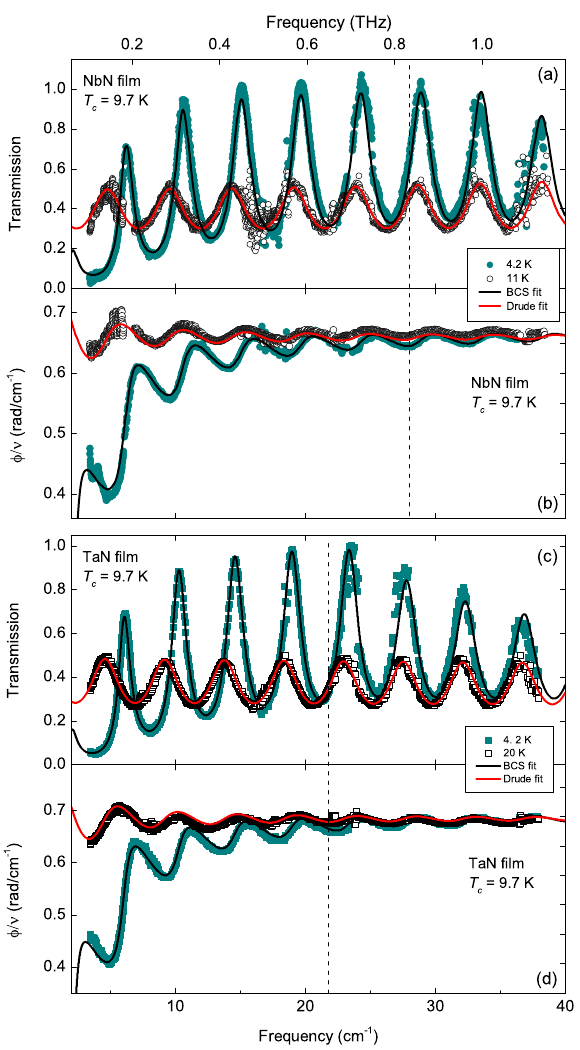}
\caption{\label{fig:NbN-raw}(Color online) Raw transmission of (a) NbN and (c) TaN and relative phaseshift of (b) NbN and (d) TaN versus frequency at two temperatures, 4.2\,K and in the normal state. Solid lines are transmission and relative phaseshift fits with the optical conductivity given by (\ref{eq:drude1}) and (\ref{eq:drude2}) in the normal state, and (\ref{eq:sigma}) at 4.2\,K. The vertical dashed lines mark the gap frequency $\nu_g=\omega_g/(2\pi c)\approx$ 28\,cm$^{-1}$ (NbN) and $\nu_g=\omega_g/(2\pi c)\approx$ 21.7\,cm$^{-1}$ (TaN).}
\par\end{centering}
\end{figure}

Comprehensive low temperature THz measurements were performed on all TaN samples using cryostat "2" exclusively. In case of the NbN samples, only measurements at 4.2\,K were carried out with "2" otherwise "1". Due to similar results we discuss only one TaN and one NbN sample on behalf of the others. Raw data of transmission and relative phaseshift versus frequency above and below $T_c$ are shown in Fig. \ref{fig:NbN-raw}. 
In general, raw data of both NbN and TaN are very similar. In the normal state, the overall Fabry-P\'erot pattern of the transmission coefficient and phaseshift does not show a frequency dependence. This indicates a constant dispersionless complex conductivity of the film, similar to those of a simple metal for frequencies much lower than the scattering rate. The transmission reaches 45 to 50\% signaling a comparably poor metallic conductivity. When the samples are cooled below $T_c$, transmission and relative phaseshift spectra change significantly. The amplitudes of the Fabry-P\'erot transmission peaks exceed the normal-state values for all studied frequencies. Starting at low frequencies, the maxima in transmission in NbN increase from 0.7 to 1 at about 24\,cm$^{-1}$ and slowly decrease to 0.9 at high frequencies. Similar behavior is apparent for TaN, however, the maxima in transmission decline more rapidly towards high frequencies. Towards low frequencies, relative phaseshift of both NbN and TaN is strongly reduced compared to the normal state. Amplitude and periodicity of $\mathcal{T}$ and $\phi/\omega$ vary with temperature and frequency signaling a considerable frequency dependence of the complex optical conductivity. 
\section{\label{sec:analysis NbN}Analysis and Discussion of N$\mathrm{b}$N}
\begin{figure}[t]
\noindent \begin{centering}
\includegraphics[scale=0.31]{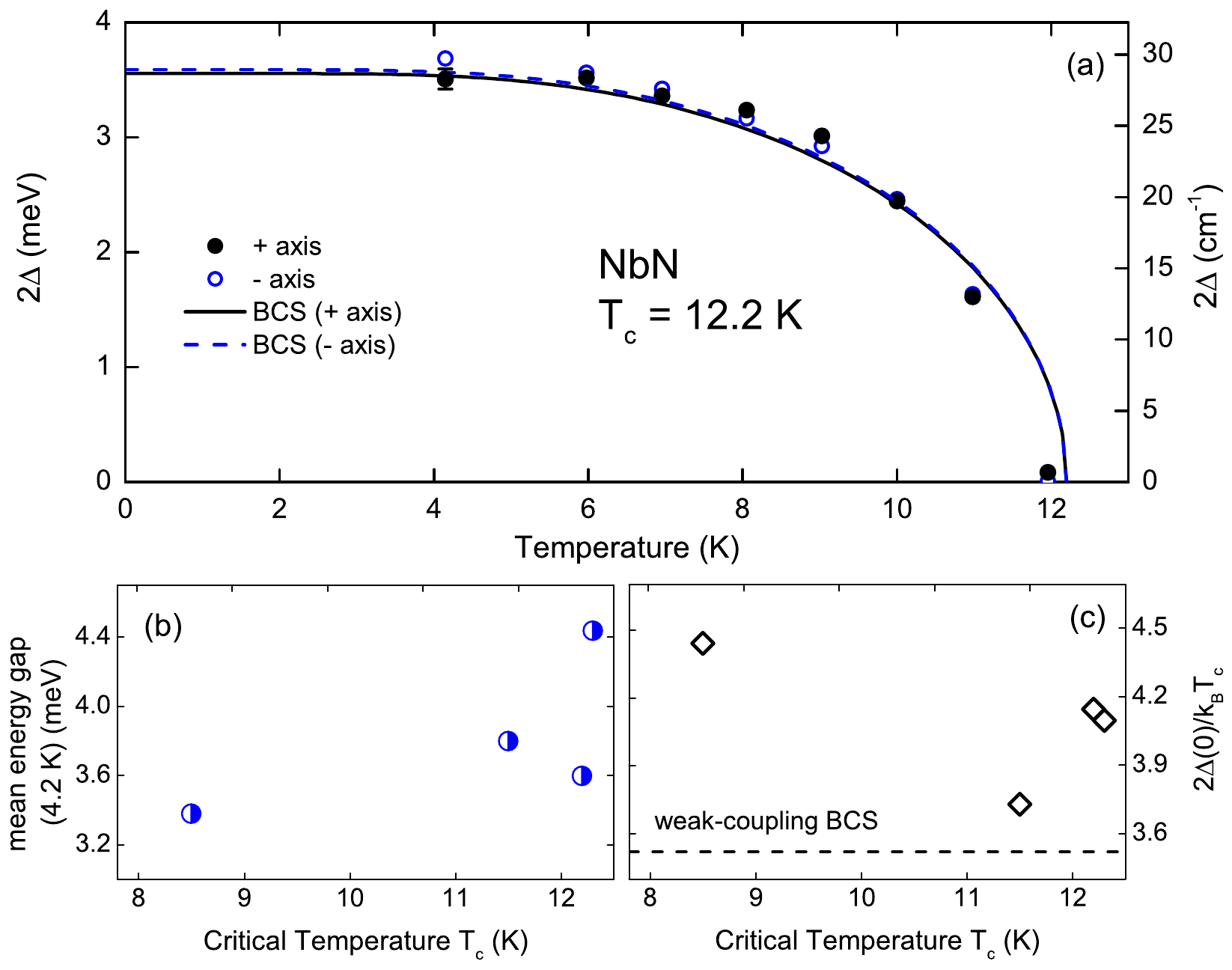}
\caption{\label{fig:NbN-gap}(Color online) (a) Energy gap $2\Delta$ versus temperature of a NbN sample for both main axes of the substrate. Solid lines are fits according to BCS theory, (\ref{eq:gap}). Results on both main axes yield very similar values of $2\Delta(T)$. The error bar shown is representative for all data. (b) $2\Delta$ at 4.2\,K and (c) ratios $2\Delta(0)/k_BT_c$ versus $T_c$ of all studied NbN films.}
\par\end{centering}
\end{figure}

All raw transmission and relative phaseshift spectra for the superconducting state measurments are fitted taking into account the frequency dependence of the conductivity given by (\ref{eq:sigma}). For each temperature, the fits for transmission and relative phaseshift are performed simultaneously with $2\Delta$ being the only fit parameter. All free-electron parameters in (\ref{eq:sigma}) (scattering rate $\nicefrac{1}{\tau}$ and dc conductivity $\sigma_{dc}$) are obtained from the normal-state measurements directly above $T_c$, which are fitted to Drude theory and taken as constant. The critical temperature $T_c$ is obtained from transport measurements, as shown in Fig. \ref{fig:rsq}. BCS fits for both transmission and relative phaseshift at 4.2\,K are shown in Fig. \ref{fig:NbN-raw}. Clearly, theory and experiment match very well for frequencies both above and below the energy gap. The same holds for the other temperatures, which allows to extract $2\Delta$ for each temperature precisely. Fig. \ref{fig:NbN-gap} displays the temperature dependent energy gap $2\Delta(T)$ of the sample with lowest disorder and highest $T_c$. 
Filled and open symbols refer to different main axes of the substrate and coincide within the range of error $\delta_{2\Delta}=$ 0.9\,meV. Data for each main axis is fitted separately with the BCS expression \cite{Ric65}
\begin{equation}
\frac {\Delta(T)}{\Delta(0)}=\tanh\left\{\frac{T_c}{T}\frac{\Delta(T)}{\Delta(0)}\right\}\label{eq:gap}.
\end{equation}
For $0.3T_c<T<0.8T_c$ the measured energy gap is slightly above the BCS curve and for $0.8T_c<T<T_c$ below. The low-temperature part displays a similar behavior as it has been observed in microwave experiments \cite{Mon13}, where 2$\Delta(T)$ does not decrease as fast as in BCS theory as $T\rightarrow T_c$. In our measurement, however, this trend is not true for temperatures close to $T_c$ and our findings do not signal a gapped normal state. Upon decreasing $T_c$, the energy gap (at 4.2\,K) is reduced from about 4.5 to 3.3\,meV, see inset of Fig. \ref{fig:NbN-gap}. Consequently, $2\Delta(T)$ depends on the amount of disorder and is greatest for low-disorder films. Furthermore, we do not see optical signatures of superconductivity in the normal state among the higher disordered samples. At 4.2\,K, 2$\Delta(T)$ is reduced as $T_c$ decreases, which underlines previous findings from tunneling studies \cite{Cho09}, see inset of Fig. \ref{fig:NbN-gap}. We calculate the ratios $2\Delta(0)/k_BT_c$ from our data assuming $2\Delta(4.2\,K) \approx 2\Delta(0)$, and find a similar anomalous behavior as in \cite{Cho09}: upon increasing disorder (decreasing $T_c$), the ratio $2\Delta(0)/k_BT_c$ is first reduced from about 4.1, indicating strong-coupling superconductivity, exhibits a minimum and becomes larger again (about 4.5) towards the highest-disorder sample studied. Our overall findings, however, favor conventional BCS superconductivity and do not feature any superconducting properties that require interpretation in terms of the SIT. This is expected to change as the film thickness is reduced.
\section{\label{analysis TaN} Analysis and Discussion on T$\mathrm{a}$N}
\begin{figure}
\noindent \begin{centering}
\includegraphics[scale=0.358]{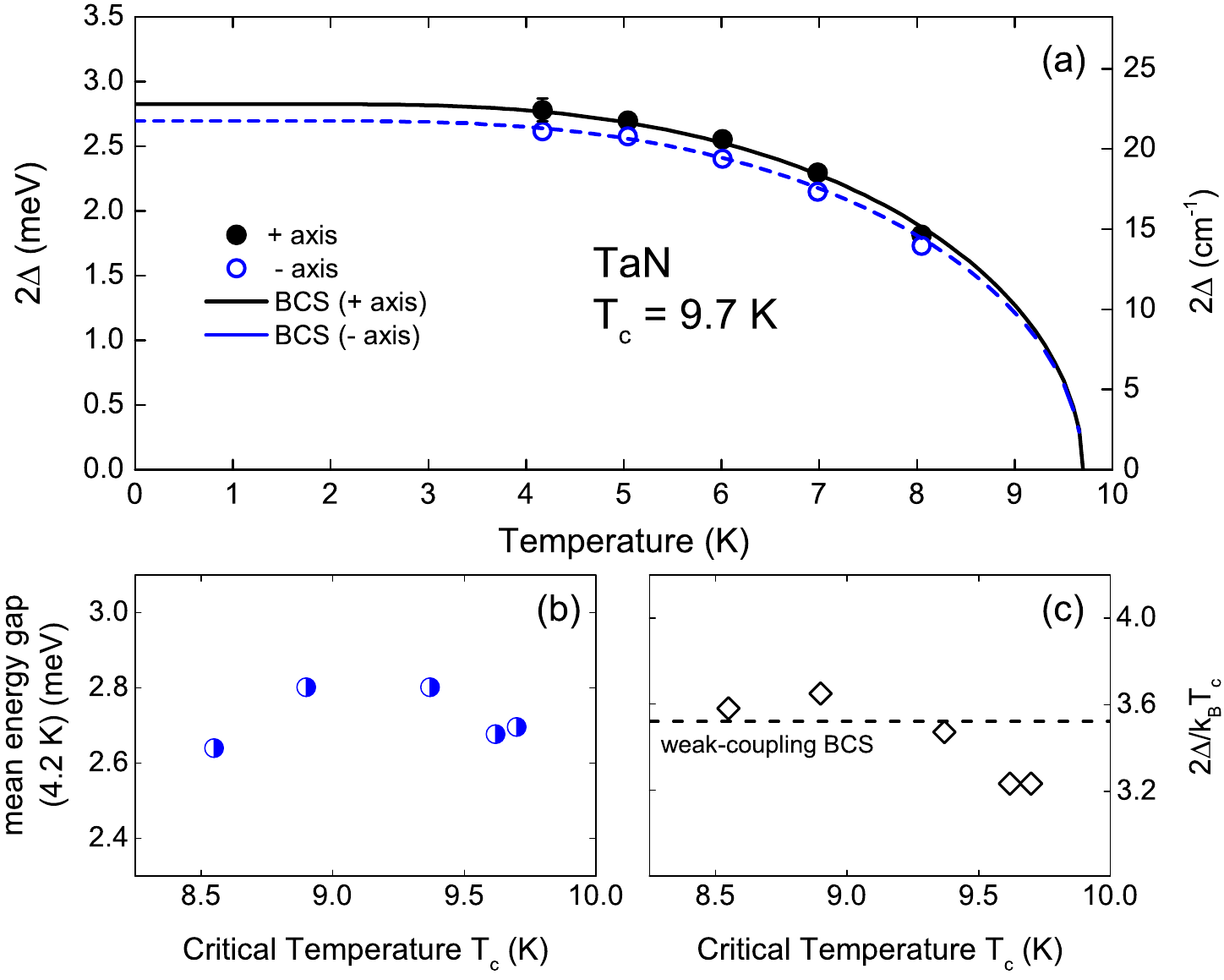}
\caption{\label{fig:TaN-gap}(Color online)(a) Energy gap $2\Delta$ versus temperature for both main axes of the substrate (denoted by "+" and "-" ). Solid lines are fits according to BCS theory, (\ref{eq:gap}). Considering each axis separately, experiment and theory match nicely. The error bar shown is representative for all data. (b) $2\Delta$ at 4.2\,K and (c) ratios $2\Delta(0)/k_BT_c$ versus $T_c$ of all studied TaN films}
\par\end{centering}
\end{figure}

\begin{figure}[H]
\noindent \begin{centering}
\includegraphics[scale=0.37]{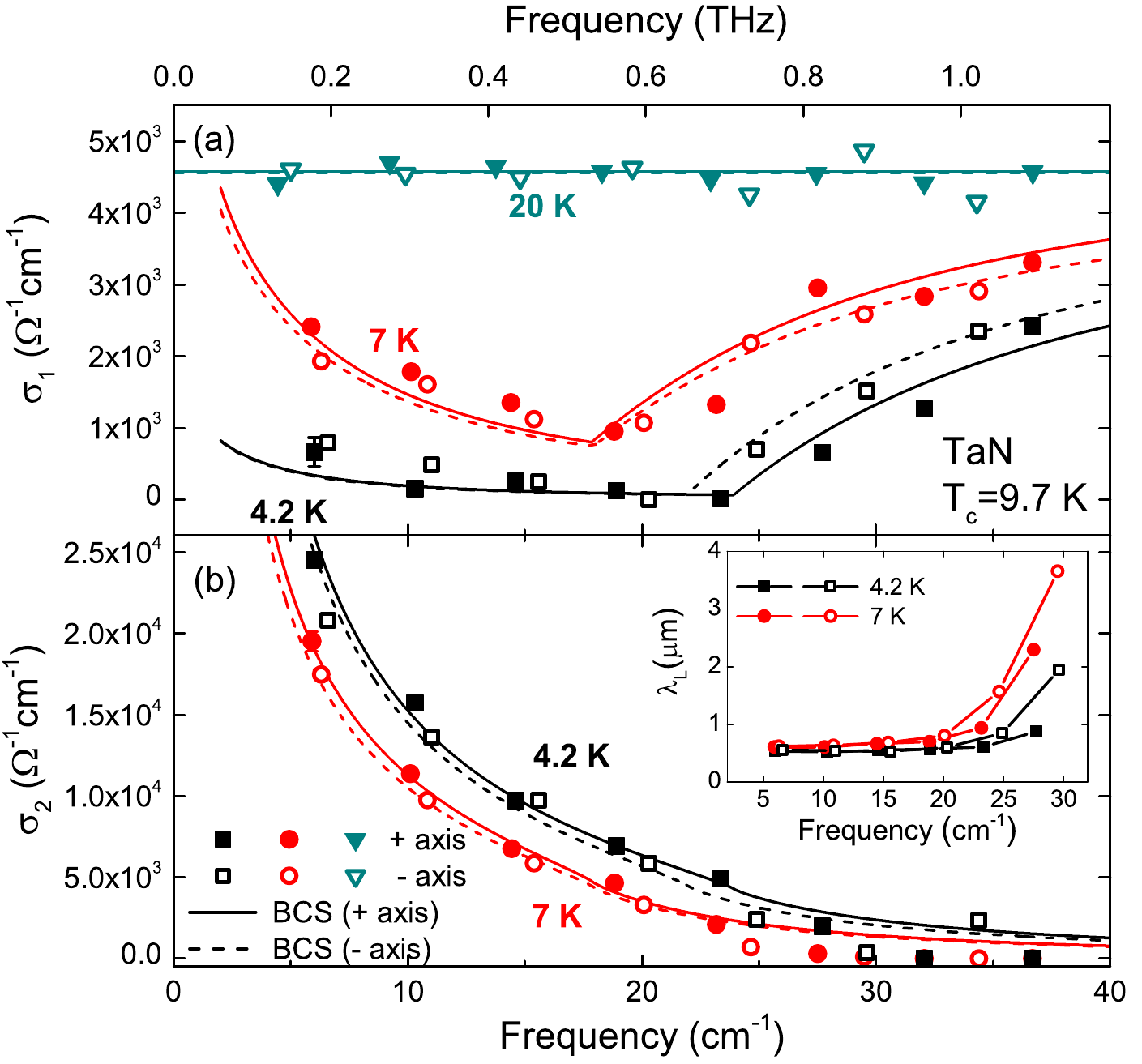}
\caption{\label{fig:TaN-sigma}(Color online) Real part (a) and imaginary part (b) of the complex conductivity versus frequency for 4.2 and 7\,K well below $T_c$ and 20\,K in the normal state for both main axes of a TaN sample. Solid and dashed lines are predictions calculated directly from raw data fits for both axes individually. The shift in frequency is caused by different peak positions due to birefringence. The kink in $\sigma_1$ signaling the energy gap is resolved clearly. The error bar shown is representative for all data. The inset shows the penetration depth as a function of frequency. $\lambda_L$ is frequency independent for $\hbar \omega < 2\Delta(T)$. }
\par\end{centering}
\end{figure}

The experimental data for TaN is processed in the same way as described for NbN. We find our experimental data in perfect agreement with weak-coupling BCS theory, as shown in Figs. \ref{fig:NbN-raw} and \ref{fig:TaN-gap} and \ref{fig:TaN-sigma} representative for all samples under study. From the BCS fits for all temperatures below $T_c$ we obtain the temperature-dependent energy gap, which is displayed in Fig. \ref{fig:TaN-gap}. 
Considering responses measured along the main axes individually, theory and experiment match nicely. Within our range of error $\delta_{2\Delta}=$ 0.9\,meV, results on both axes (denoted by "+" and "-" ) are the same. We attribute the fairly constant difference for both axes to a small systematic experimental error aligning the substrate\'{}s axes and the THz polarization. The inset of Fig. \ref{fig:TaN-gap} displays $2\Delta$ at 4.2\,K for all studied samples versus $T_c$. While $2\Delta$ tends to decrease as $T_c$ is reduced in the case of NbN, we do not see a significant dependence for TaN.
Since the ratio between highest and lowest critical temperatures $T_c^{max}/T_c^{min}$ is 1.13 and 1.44 for TaN and NbN, respectively, we can only access a smaller range here which makes it difficult to establish similar conclusions from our TaN experiments. We calculate the ratio $2\Delta(0)/k_BT_c$ from our experimental data to be closely scattered around the weak-coupling prediction 3.52 \cite{Bar57}, which underlines pure weak-coupling BCS superconductivity in this range of disorder and critical temperatures.

We also analyze the raw data with single-peak fits to extract the frequency dependence of the complex conductivity $\hat{\sigma}(\omega)=\sigma_1(\omega)+\mathrm{i}\sigma_2(\omega)$ independently from a predefined model. For this, a narrow frequency range of 0.5 to 1\,cm$^{-1}$ is chosen for fitting each single Fabry-P\'erot peak separately with $\epsilon_1$ and $\epsilon_2$ as fitting parameters. Subsequently, $\sigma_1$ and $\sigma_2$ are calculated via $\hat{\epsilon}=\epsilon_{\infty}+\mathrm{i}4\pi \hat{\sigma}/\omega$ for each peak frequency with $\epsilon_{\infty}=1$. Fig. \ref{fig:TaN-sigma} displays the conductivity for 4.2 and 7 K well below $T_c=9.7$ K and at 20 K in the normal state. Closed and open symbols are for the different main axes. Data points corresponding to different main axes are shifted in frequency. The reason is that $\epsilon_1$ of the birefringent substrate is slightly different for both axes leading to a shift in peak position. The solid lines are BCS predictions based on the raw-data fits. Within the range of error, theory and experiment (for both main axes) match reasonably. $2\Delta(T)$ is signaled by the kink in $\sigma_1$ and is clearly resolved within the experimental frequency range. At 20 K, $\sigma_1$ does not depend on the frequency and corresponds to the constant value $\sigma_1=\sigma_{dc}=4580$ $\Omega^{-1}$cm$^{-1}$ extracted from Drude fits. The studied frequency range is well below the scattering rate $\nicefrac{1}{\tau}$ and, thus, $\sigma_2$ is immeasurably small in the normal state at 20 K. In the superconducting state, however, $\sigma_2$ rises notably and approaches the zero-temperature prediction $\sigma_2(T=0)\propto\omega^{-1}$ for $\hbar \omega<2\Delta(T)$ , which is consistent with the $\delta$-function-like conductivity $\sigma_1$ at zero frequency. This trend in $\sigma_2$ is also seen in the superconducting penetration depth $\lambda_L$. 
The inset of Fig. \ref{fig:TaN-sigma}(b) displays $\lambda_L$ versus frequency for both main axes at 4.2 and 7 K calculated via \cite{dre02}
\begin{equation}
\lambda_L=\sqrt{\frac{c^2}{4\pi\omega\sigma_2}}\label{eq:pendepth}.
\end{equation}
For frequencies $\hbar \omega<2\Delta(T)$ and a $\delta$-function-like $\sigma_1$, $\lambda_L$ does not significantly depend on frequency. Despite the finite temperatures, such frequency-independent behavior is supported by the data in Fig. \ref{fig:TaN-sigma}. At 4.2\,K, $\lambda_L$ remains fairly constant for frequencies lower than 24\,cm$^{-1}$ and approaches $\sim 500$ nm for $\omega\rightarrow 0$. At 7 K, similar behavior is apparent for frequencies lower than 15\,cm$^{-1}$, and $\lambda_L$ approaches $\sim 600$ nm. The zero-frequency values of $\lambda_L$ are very similar to ones obtained from previous transport measurements \cite{Eng12b}. By and large, we find the charge carrier dynamics of our TaN films to be perfectly described within weak-coupling BCS theory.
\section{Summary}
We have described the performance and characteristics of our frequency-domain THz spectroscopic experimental techniques. Furthermore, we give a comprehensive overview of the analyses we employ and the theoretical foundation on which they are based. We studied charge carrier dynamics and superconducting properties of NbN and TaN thin films with different values of $T_c$ at low temperatures above and below $T_c$ in the THz frequency range from 0.09 to 1.1\,THz (3 to 38\,cm$^{-1}$). Coherent and continuous THz radiation was generated by frequency-tunable high power backward wave oscillators covering the frequency range below and above the energy gap $2\Delta(T)$. We fit our experimental data of transmission and phaseshift to a combination of Fresnel equations, BCS- and Drude theory and describe superconducting properties and charge carrier dynamics. The results for both NbN and TaN are very well described within the framework of BCS theory. The agreement between our results and previous measurements on NbN from another group confirm that our experimental approach is perfectly suited to study correlated electron phenomena in thin film systems. The dependence of the ratio $2\Delta(0)/k_{B}T_c$ and $2\Delta$ on $T_c$ we obtain for our NbN films resemble anomalous behavior as it has been reported previously, whereas the ratios for our TaN films do not significantly deviate from the weak-coupling BCS prediction. Absolute values of $2\Delta$ and the ratio $2\Delta(0)/k_BT_c$ for NbN are in good agreement with previous publications, and thus underline the reliability and accuracy of our experiment and analyses. This also endorses our results on TaN, a compound that so far has not been studied in detail concerning THz charge carrier dynamics.
\section{Acknowledgment}
We would like to thank Sina Zapf for fruitful discussion, Qiao Guo for help preparing the TaN samples, Elena Zhukova for critically reading the manuscript, and the DFG for financial support. This study was partly financially supported by the Russian Foundation for Basic Research (project11-02-00199-a).


\begin{IEEEbiography}[{\includegraphics[width=1in,height=1.25in,clip,keepaspectratio]{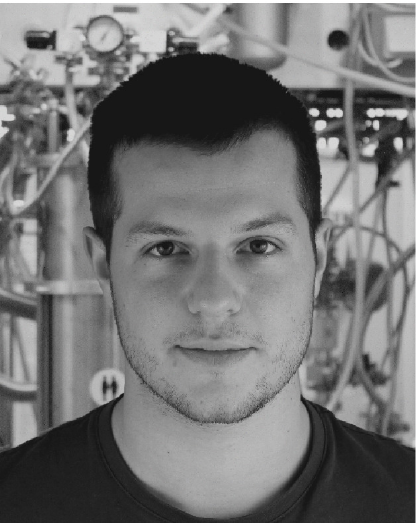}}]{Uwe S. Pracht}
received his B.Sc. and M.Sc. degrees in physics from the Universit\"at Stuttgart, Germany, in 2010 and 2012, respectively. He is currently working towards his Ph.D. degree in physics from the Universit\"at Stuttgart, Germany. His current research activities focus on THz spectroscopy at $^4$He- and $^3$He temperatures on compounds relevant for superconductor-insulator quantum phase transition.
\end{IEEEbiography}
\begin{IEEEbiography}[{\includegraphics[width=1in,height=1.25in,clip,keepaspectratio]{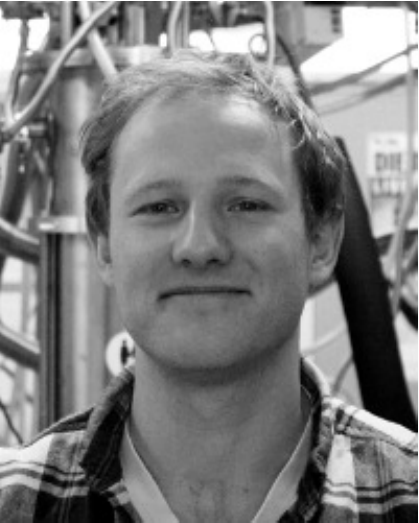}}]{Eric Heintze}
received his diploma degree in physics from the Universit\"at Stuttgart, Germany, in 2010. He is currently working towards his Ph.D. degree in physics from the Universit\"at Stuttgart, Germany. His research activities focus on pulsed ESR, W-band and molecular magnets.
\end{IEEEbiography}
\begin{IEEEbiography}[{\includegraphics[width=1in,height=1.25in,clip,keepaspectratio]{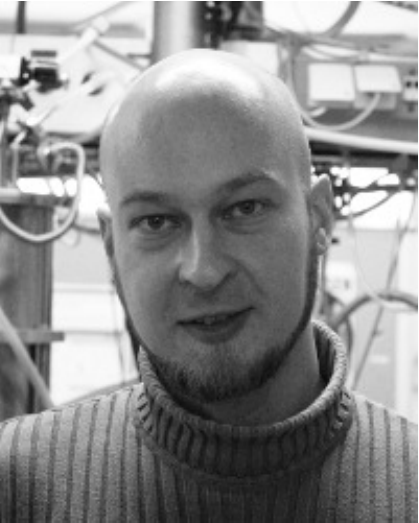}}]{Conrad Clauss}
received his diploma degree in physics from the Universit\"at Stuttgart, Germany, in 2009. He is currently working towards his Ph.D. degree in physics from the Universit\"at Stuttgart, Germany. His research activities focus on quantum optics with molecular magnets.
\end{IEEEbiography}
\begin{IEEEbiography}[{\includegraphics[width=1in,height=1.25in,clip,keepaspectratio]{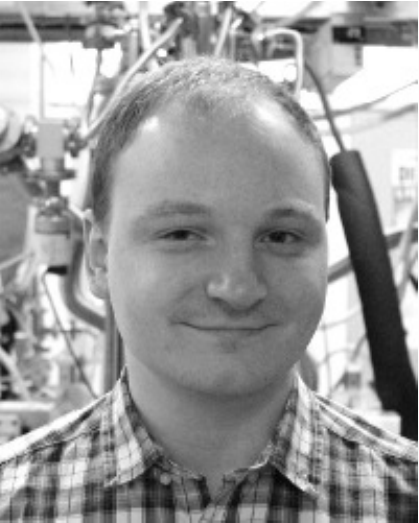}}]{Daniel Hafner}
received his B.Sc. and M.Sc. degrees in physics from the Universit\"at Stuttgart, Germany, in 2010 and 2012, respectively. He is currently working towards his Ph.D. degree in physics from the Universit\"at Stuttgart, Germany. His research activities focus on microwave strip line resonator experiments on correlated electron systems.
\end{IEEEbiography}
\begin{IEEEbiography}[{\includegraphics[width=1in,height=1.25in,clip,keepaspectratio]{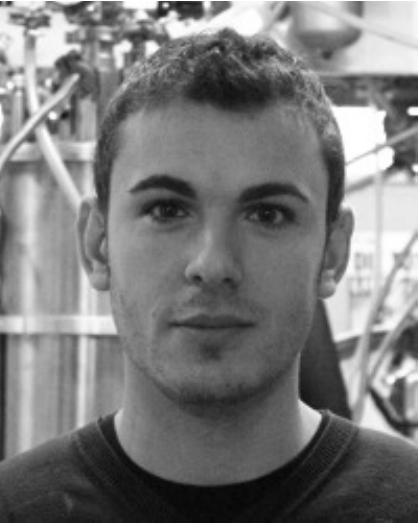}}]{Roman Bek}
received his B.Sc. and M.Sc. degrees in physics from the Universit\"at Konstanz, Germany, in 2010, and the Universit\"at Stuttgart, Germany, in 2012, respectively. He is currently working towards his Ph.D. degree in physics from the Universit\"at Stuttgart, Germany. His research activities focus on vertical-external-cavity-surface-emitting-lasers (VECSEL).
\end{IEEEbiography}
\begin{IEEEbiography}[{\includegraphics[width=1in,height=1.25in,clip,keepaspectratio]{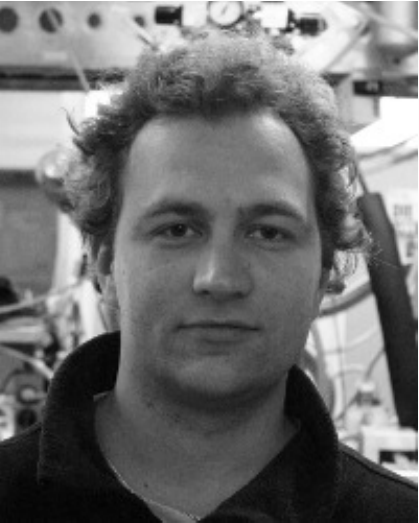}}]{Sergey Gelhorn}
received his B.Sc. degree in physics from the University of Stuttgart, in 2012. He is currently working towards his M.Sc. degree in physics from the Universit\"at Stuttgart, Germany.
\end{IEEEbiography}
\begin{IEEEbiographynophoto}{David Werner}
is currently working towards his diploma degree in physics from the University of Stuttgart.
\end{IEEEbiographynophoto}
\begin{IEEEbiography}[{\includegraphics[width=1in,height=1.25in,clip,keepaspectratio]{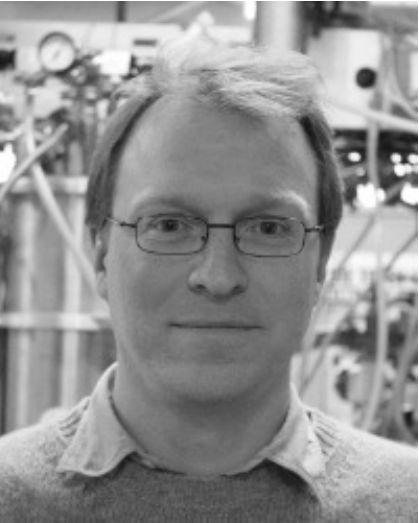}}]{Marc Scheffler}
studied physics at the Technische Universit\"at Braunschweig,
Germany, and he received the M.Sc. degree from the University of Maryland,
College Park in 1998, and the PhD degree from the University of Stuttgart,
Germany in 2004. After a postdoc period at the Delft University of
Technology, the Netherlands, he now holds a permanent position at the
University of Stuttgart. His research activities address the electronic
properties of correlated metals and superconductors, which he studies with
microwave and THz spectroscopy.
\end{IEEEbiography}
\begin{IEEEbiography}[{\includegraphics[width=1in,height=1.25in,clip,keepaspectratio]{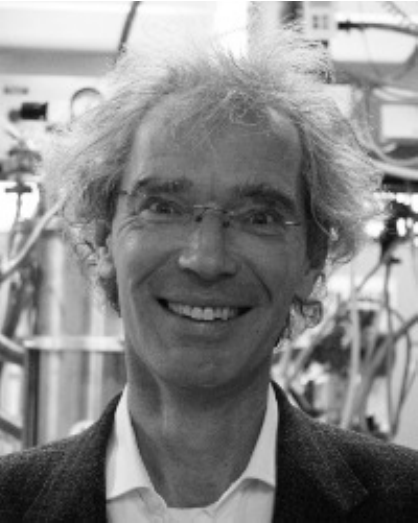}}]{Martin Dressel}
received his Ph.D. degree from the Universit\"at G\"ottingen, Germany, in 1989, for studying magnetic-transport in organic conductors and superconductors by microwave. Since 1989, he has held positions at the Laser-Laboratory G\"ottingen, Germany, the University of British Columbia, Vancouver, and the University of California, Los Angeles. He received his Habilitation from the technical Universit\"at Darmstadt, Germany, in 1996, and then, joined the Center of Electronic Correlations and Magnetism, Universit\"at Ausgburg, Germany. Since 1998 he is Head of the 1. Physikalisches Institut, Universit\"at Stuttgart, Germany, and currently Dean of the Faculty. His research interests include the electronic and magnetic properties
of low-dimensional electron systems. The group is renowned for its investigations on the electrodynamics of electronically correlated matter.
\end{IEEEbiography}
\begin{IEEEbiography}[{\includegraphics[width=1in,height=1.25in,clip,keepaspectratio]{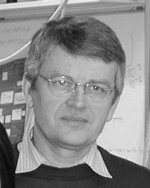}}]{Boris Gorshunov}received his B.Sc. degree in physics from the Moscow Physical-Engineering Institute, Moscow, Russia, in 1978, and the Ph.D. degree from the Russian Academy of Sciences, Moscow, in 1988 for his work in submillimeter spectroscopy of solids. He was with the Lebedev Physical Institute, Moscow, and now at the A.M. Prokhorov Institute of General Physics, Russian Academy of Sciences, where he is currently Head of a laboratory. He spent some years abroad to work at universities in Regensburg, Los Angeles, Stuttgart, and other laboratories. His current research interests include the field of strongly correlated electronic phenomena in solids.
\end{IEEEbiography}
\begin{IEEEbiography}[{\includegraphics[width=1in,height=1.25in,clip,keepaspectratio]{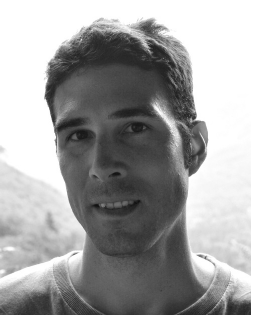}}]{Daniel Sherman}
received his education in Physics from Bar-Ilan University. Finished his M.Sc.
degree in 2010 and is currently studying towards a PhD degree.
For the first two years of his Ph.D. he studied the disordered tuned
superconductor to insulator transition. The main focus has been DC tunneling
and transport measurements. Most recently he has been working on a new
approach to investigate the disorder effects on superconductors using far
infra-red spectroscopy.
\end{IEEEbiography}
\begin{IEEEbiography}[{\includegraphics[width=1in,height=1.25in,clip,keepaspectratio]{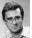}}]{Konstantin Il\'{}in}
received the Ph.D. degree in solid-state physics from Moscow State Pedagogical University (MSPU), Moscow, Russia, in 1998. From 1997 to 1998, he was a Visiting Scientist with the Electrical and Computer Engineering Department, University of Massachusetts at Amherst, and with the Electrical Engineering Department, University of Rochester, Rochester, NY. From January 1998 to June 1999, he was an Assistant Professor with the Physics Department, MSPU. From 1999 to 2003, he was a Scientific Researcher with the Institute of Thin Films and Interfaces, Research Center Juelich, Juelich, Germany. In June 2003, he joined the Institute of Micro- and Nano-electronic Systems, University of Karlsruhe, Karlsruhe, Germany, where he currently develops technology of ultrathin films of conventional superconductors for receivers of electromagnetic radiation. His research interests include fabrication and study of normal state and superconducting properties of submicrometer- and nanometer-sized structures from ultrathin films of disordered superconductors. 
\end{IEEEbiography}
\begin{IEEEbiography}[{\includegraphics[width=1in,height=1.25in,clip,keepaspectratio]{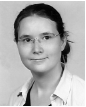}}]{Dagmar Henrich}
recieved her diploma in Physics from the University of Karlsruhe (TH), Germany, in 2007. She then joined the Institute of Micro- and Nanoelectronic Systems at the Karlsruhe Institute of Technology, where she is involved in the development of superconducting thin films and single photon detectors.
\end{IEEEbiography}

\begin{IEEEbiography}[{\includegraphics[width=1in,height=1.25in,clip,keepaspectratio]{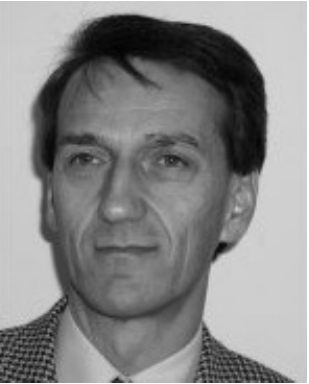}}]{Michael Siegel}
received the Diploma degree in physics and the Ph.D. degree in solid state physics from the Moscow State University, Moscow, U.S.S.R., in 1978 and 1981, respectively. In 1981, he joined the University of Jena where he held positions as Staff Member and later as Group Leader in the Superconductive Electronic Sensor Department. His research was oriented on non-linear superconductor-semiconductor devices for electronic applications. In 1987, he initiated research at the University of Jena in thin-film high temperature superconductivity (HTS) for Josephson junction devices, mainly for SQUID. In 1991, he left to join the Institute for Thin Film and Ion Technology at Research Center Juelich. There he worked on development and application of HTS Josephson junctions, SQUID, microwave arrays and mixers, and high-speed digital circuits based on rapid single- flux-quantum logic. In 2002, he received a Full Professor position at University of Karlsruhe, Germany, where he is now the Director of the Institute of Micro- and Nanoelectronic Systems. His research includes transport phenomena in superconducting, quantum and spin dependent tunnelling devices. He has authored over 200 technical papers.\end{IEEEbiography}

\end{document}